\documentclass[lettersize,journal]{IEEEtran}
\usepackage{amsmath,amssymb,amsfonts}
\usepackage{algorithmic}
\usepackage{graphicx}
\usepackage{textcomp}
\usepackage{xcolor}
\usepackage{subfigure}

\usepackage{bm}
\usepackage{algorithm}
\usepackage{algorithmic}
\usepackage{soul}
\usepackage{balance}
\usepackage{cite}
\usepackage[version=3]{acro}

\usepackage{url}
\usepackage{verbatim}
\usepackage{acro} 
\usepackage{enumitem}

\DeclareAcronym{ICI}{short = ICI ,long  = inter-carrier interference ,tag = abbrev}
\DeclareAcronym{ISI}{short = ISI ,long  = inter-symbol interference ,tag = abbrev}
\DeclareAcronym{OFDM}{short = OFDM ,long  = orthogonal frequency division multiplexing ,tag = abbrev}
\DeclareAcronym{OTFS}{short = OTFS ,long  = orthogonal time frequency space ,tag = abbrev}
\DeclareAcronym{DFT}{short = DFT ,long  = discrete Fourier transform ,tag = abbrev}
\DeclareAcronym{IDFT}{short = IDFT ,long  = inverse discrete Fourier transform ,tag = abbrev}
\DeclareAcronym{ISFFT}{short = ISFFT ,long  = inverse symplectic fast Fourier transform ,tag = abbrev}
\DeclareAcronym{SFFT}{short = SFFT ,long  = symplectic fast Fourier transform ,tag = abbrev}
\DeclareAcronym{PAPR}{short = PAPR ,long  = peak-to-average power ratio ,tag = abbrev}
\DeclareAcronym{6G}{short = 6G ,long  = sixth generation ,tag = abbrev}
\DeclareAcronym{5G}{short = 5G ,long  = fifth generation ,tag = abbrev}
\DeclareAcronym{4G}{short = 4G ,long  = fourth generation ,tag = abbrev}
\DeclareAcronym{DD}{short = DD ,long  = delay-Doppler ,tag = abbrev}
\DeclareAcronym{TF}{short = TF ,long  = time-frequency ,tag = abbrev}
\DeclareAcronym{MIMO}{short = MIMO ,long  = multiple input multiple output ,tag = abbrev}
\DeclareAcronym{3GPP}{short = 3GPP ,long  = 3rd generation partnership program ,tag = abbrev}
\DeclareAcronym{MA}{short = MA ,long  = multiple access ,tag = abbrev}
\DeclareAcronym{MUI}{short = MUI ,long  = multi-user interference ,tag = abbrev}
\DeclareAcronym{SINR}{short = SINR ,long  = signal to interference plus noise ratio ,tag = abbrev}
\DeclareAcronym{TDMA}{short = TDMA ,long  = time-division multiple access ,tag = abbrev}
\DeclareAcronym{FDMA}{short = FDMA ,long  = frequency-division multiple access ,tag = abbrev}
\DeclareAcronym{SNR}{short = SNR ,long  = signal to noise ratio ,tag = abbrev}
\DeclareAcronym{FCP}{short = FCP ,long  = full-CP,tag = abbrev}
\DeclareAcronym{HM}{short = HM ,long  = high mobility,tag = abbrev}
\DeclareAcronym{LM}{short = LM ,long  = low mobility,tag = abbrev}

\DeclareAcronym{AWGN}{short = AWGN ,long  =  additive white Gaussian noise,tag = abbrev}

\DeclareAcronym{BS}{short = BS ,long  =  base station,tag = abbrev}

\DeclareAcronym{DL}{short = DL ,long  =  downlink ,tag = abbrev}

\DeclareAcronym{UL}{short = UL ,long  =  uplink,tag = abbrev}
\DeclareAcronym{UE}{short = UE ,long  =  user equipment,tag = abbrev}
\DeclareAcronym{BER}{short = BER ,long  =  bit-error rate,tag = abbrev}

\DeclareAcronym{SC}{short = SC ,long  =  single-carrier,tag = abbrev}

\DeclareAcronym{BW}{short = BW ,long  =  band-width,tag = abbrev}

\DeclareAcronym{MSE}{short = MSE ,long  =  mean-square-error,tag = abbrev}

\DeclareAcronym{MU}{short = MU ,long  =  multi-user,tag = abbrev}

\DeclareAcronym{TD}{short = TD ,long  =  time delay,tag = abbrev}

\DeclareAcronym{ST}{short = ST ,long  =  symbol time,tag = abbrev}

\DeclareAcronym{SCS}{short = SCS ,long  =  subcarrier spacing,tag = abbrev}

\DeclareAcronym{PEP}{short = PEP ,long  =  pairwise-error probabilities,tag = abbrev}
\DeclareAcronym{BPSK}{short = BPSK ,long  =  binary phase shift keying,tag = abbrev}
\DeclareAcronym{QAM}{short = QAM ,long  =  quadrature amplitude modulation,tag = abbrev}
\DeclareAcronym{QPSK}{short = QPSK ,long  =  quadrature phase shift keying,tag = abbrev}
\DeclareAcronym{CP}{short = CP ,long  =  cyclic-prefix,tag = abbrev}

\DeclareAcronym{SDR}{short = SDR ,long  =  software-defined radio,tag = abbrev}

\DeclareAcronym{NTN}{short = NTN ,long  =  non-terrestrial networks,tag = abbrev}

\DeclareAcronym{mMIMO}{short = mMIMO ,long  =  massive multiple input multiple output,tag = abbrev}

\DeclareAcronym{SIC}{short = SIC ,long  =  successive interference cancellation,tag = abbrev}

\DeclareAcronym{RIS}{short = RIS ,long  =  reconfigurable intelligent surfaces,tag = abbrev}

\DeclareAcronym{NOMA}{short = NOMA ,long  =  non orthogonal MA,tag = abbrev}

\DeclareAcronym{RSMA}{short = RSMA ,long  =  rate splitting MA,tag = abbrev}

\DeclareAcronym{TA}{short = TA ,long  =  timing advance, tag = abbrev}

\DeclareAcronym{FDE}{short = FDE ,long  =  frequency-domain equalization , tag = abbrev}

\DeclareAcronym{AFDM}{short = AFDM ,long  =  affine frequency division multiplexing , tag = abbrev}

\DeclareAcronym{OCDM}{short = OCDM ,long  =  orthogonal chirp division multiplexing, tag = abbrev}

\DeclareAcronym{OTSM}{short = OTSM ,long  =  orthogonal time sequency multiplexing, tag = abbrev}

\DeclareAcronym{ODDM}{short = ODDM ,long  =  orthogonal delay Doppler division multiplexing, tag = abbrev}

\DeclareAcronym{MMDS}{short = MMDS ,long  =  misaligned maximum excess delay , tag = abbrev}

\DeclareAcronym{AMDS}{short = AMDS ,long  =  aligned maximum delay spread, tag = abbrev}

\DeclareAcronym{SE}{short = SE ,long  =  spectral efficiency, tag = abbrev}

\DeclareAcronym{Rx}{short = Rx ,long  =  receiver, tag = abbrev}

\DeclareAcronym{Tx}{short = Tx ,long  =  transmitter, tag = abbrev}

\DeclareAcronym{PHY}{short = PHY ,long  =  physical layer, tag = abbrev}

\DeclareAcronym{ZC}{short = ZC ,long  =  Zadoff-Chu, tag = abbrev}

\DeclareAcronym{TO}{short = TO ,long  =  time-offset, tag = abbrev}
\DeclareAcronym{GB}{short = GB ,long  =  guard band, tag = abbrev}
\DeclareAcronym{DG}{short = DG ,long  =  delay guard, tag = abbrev}

\DeclareAcronym{OFDMA}{short = OFDMA ,long  =  OFDM multiple-access, tag = abbrev}

\DeclareAcronym{IOR}{short = IOR ,long  =  input--output relationship, tag = abbrev}

\DeclareAcronym{ZT}{short = ZT ,long  =  Zak transform, tag = abbrev}

\DeclareAcronym{IZT}{short = IZT ,long  =  inverse Zak transform, tag = abbrev}

\DeclareAcronym{DaFT}{short = DaFT ,long  =  discrete affine Fourier transform, tag = abbrev}

\DeclareAcronym{IDaFT}{short = IDaFT ,long  =  inverse discrete affine Fourier transform, tag = abbrev}

\DeclareAcronym{DFnT}{short = DFnT ,long  =  discrete Fresnel transform, tag = abbrev}

\DeclareAcronym{IDFnT}{short = IDFnT ,long  =  inverse discrete Fresnel transform, tag = abbrev}

\DeclareAcronym{RC}{short = RC ,long  = raised cosine, tag = abbrev}

\DeclareAcronym{DeDMA}{short = DeDMA ,long  = delay division multiple access, tag = abbrev}

\DeclareAcronym{DoDMA}{short = DoDMA ,long  = Doppler division multiple access, tag = abbrev}

\DeclareAcronym{SeqDMA}{short = SeqDMA ,long  = sequency division multiple access, tag = abbrev}

\DeclareAcronym{TM}{short = TM ,long  = time misalignment, tag = abbrev}

\DeclareAcronym{mmWave}{short = mmWave ,long  = millimeter-wave, tag = abbrev}

\DeclareAcronym{CCDF}{short = CCDF ,long  = complementary cumulative distribution function, tag = abbrev}

\begin{document}

\title{Timing Misalignment and Its Impact on OTFS and Other Delay-Based Waveforms}
  \author{ Badr Eddine Ouakouak, Salah Eddine Zegrar,~\IEEEmembership{Member,~IEEE}, and H\"{u}seyin Arslan,~\IEEEmembership{Fellow,~IEEE}

\thanks{Badr Eddine Ouakouak and H\"{u}seyin Arslan are with the Department of Electrical and Electronics Engineering, Istanbul Medipol University, Istanbul, 34810, Turkey (e-mail: badr.ouakouak@std.medipol.edu.tr; huseyinarslan@medipol.edu.tr).}
\thanks{Salah Eddine Zegrar is with KAUST, Thuwal, 23955, Saudi Arabia (e-mail: salaheddine.zegrar@kaust.edu.sa).}}


\maketitle
\begin{abstract}
The shift towards sixth-generation (6G) wireless networks calls for advanced waveforms to support emerging applications such as ultra-reliable and massive machine-type communications. Among the proposed candidates, orthogonal time frequency space (OTFS), orthogonal delay Doppler division multiplexing (ODDM), orthogonal time sequency multiplexing (OTSM), and orthogonal chirp division multiplexing (OCDM) share the common feature of exploiting the delay domain for data transmission. While these waveforms offer significant advantages in high-mobility and sensing scenarios, their integration into multi-user uplink systems reveals a critical challenge: timing misalignment (TM). Unlike OFDM, where residual timing offsets translate into phase rotations, delay-based waveforms experience actual shifts of resources, whose impact we investigate in this work. We show that the aforementioned waveforms can be unified under a time-invariant setup through a common input--output relation, which enables a unified analysis of TM in uplink resource allocation. The results reveal a fundamental trade-off between interference mitigation and guard overhead in the presence of TM. We further benchmark the observed behavior against OFDM under Doppler spread as a reference impairment and highlight the analogy between TM in delay-based allocation and Doppler spread in frequency-based allocation. The results show that delay-based waveforms can maintain a net performance advantage over OFDM in the considered setup. Beyond this unified Doppler-free analysis, we evaluate the joint impact of Doppler spread and TM on each waveform individually. We also investigate BER performance and show how coding can mitigate the resulting interference. This study therefore establishes a problem framework that motivates the development of future compensation and scheduling strategies for practical multi-user operation of delay-based waveforms.
\end{abstract}


\section{Introduction} 
The rapid evolution of wireless communication systems has led to unprecedented growth in mobile data traffic, the proliferation of connected devices, and the emergence of new applications requiring ultra-reliable and low-latency communication. The vision for \ac{6G} encompasses not only higher data rates and lower latency but also enhanced capabilities such as high-mobility communication, ultra-massive machine-type communication, and extremely low-power communication \cite{zhang20196g}. Achieving these targets requires innovations across all layers of the network, with the physical layer and waveform design being critical areas of focus \cite{10382693}.

\Ac{OFDM} has been the cornerstone waveform for \ac{5G} due to its robustness against multipath fading, high \ac{SE}, and ease of implementation. One of the key reasons for the success of OFDM so far is its ability to support diverse requirements needed for full network operation. This includes OFDM's inherent fit for \ac{MIMO} systems and its suitability for \ac{MU} systems, such as \ac{OFDMA}. On top of that, with the help of \ac{CP}, OFDM has an advantage in \ac{UL} \ac{MU} reception. The \ac{BS} does not need to know the exact point at which each \ac{UE}'s signal arrives. With CP-OFDM, the signal from the UE can still arrive with a \ac{TM} with respect to the reference \ac{UL} time at the BS, provided that this does not push the channel's delay spread beyond the CP duration. This particular issue causes no major problems related to \ac{MUI} or \ac{SE} \cite{park2003performance}. This TM manifests only as a frequency-dependent phase rotation. Therefore, the resulting effective \ac{TF} channel neither introduces additional \ac{ISI} nor \ac{ICI}. This facilitates a flexible and spectrally efficient \ac{MA} application. On the other hand, since the phase rotation is frequency-selective (although amplitude remains unchanged), it reduces coherence between subcarriers, increasing the channel's frequency selectivity in terms of phase. Such selectivity is managed by adopting high pilot densities (e.g., \(1/2\)), as is typically employed in practical 5G NR deployments \cite{3gpp38.211}. 

% OFDM Cannot Fulfill All Requirements
\textcolor{black}{Nevertheless, as the demands for \ac{6G} systems become more diverse and stringent \cite{andrews20246}, the limitations of OFDMA become more pronounced in channels with strong time variation. In high-mobility scenarios, the wireless channel is characterized not only by multipath delay spread but also by Doppler shifts and Doppler spread. These Doppler components reduce the channel coherence time and make the channel vary within an OFDM symbol or across adjacent OFDM symbols. As a result, the orthogonality among subcarriers is no longer perfectly preserved, leading to \ac{ICI}, degraded channel estimation accuracy, and increased pilot and equalization overhead. In multi-user uplink transmission, the problem becomes more involved because different users may experience different delay-Doppler characteristics depending on their velocities, propagation directions, and scattering environments. These channel-induced impairments motivate the investigation of alternative waveforms that provide more structured representations of dispersive channels and offer additional flexibility for resource allocation and interference management.}

% Different Candidate Waveforms for 6G
To address the limitations of OFDM, several candidate waveforms have been proposed, each offering many advantages and greater flexibility. These waveforms include \ac{OTFS}, \ac{ODDM}, \ac{OTSM}, and \ac{OCDM}, which were introduced respectively in \cite{hadani2017orthogonal}, \cite{lin2022orthogonal}, \cite{thaj2021orthogonal}, and \cite{ouyang2016orthogonal}.
%%%%%%%%%%%%%%%%%%%%%% 

A common characteristic among these waveforms \cite{hadani2017orthogonal,lin2022orthogonal,thaj2021orthogonal,ouyang2016orthogonal}, as will be detailed later in this work, is the utilization of the delay domain (or its equivalent) for data symbol modulation. As a result, they are collectively categorized as \textit{delay-based waveforms}. This gives them more control and flexibility, compared to OFDM, to handle a wider range of channel conditions for both sensing and communication \cite{10445333,10772237}, and capture its full diversity \cite{surabhi2019diversity}. Hence, they have achieved remarkable success in improving on OFDM and mitigating its limitations.  However, determining whether a given modulation technique can serve as an alternative to OFDM or complement it in addressing 6G requirements necessitates a comprehensive assessment of its practical network performance. Key considerations include its suitability for MU operation, support for \ac{MA} schemes, and compatibility with existing network architectures. In this work, we specifically focus on enabling MU operation using these waveforms.

\textcolor{black}{In practical MU operation, acquiring reliable channel information is a prerequisite for designing efficient MA and resource allocation schemes. The BS must obtain, either through uplink pilots, downlink feedback, or waveform-specific channel-estimation procedures, sufficient information about each UE's effective channel support, delay spread, and possible residual timing offset. This information is essential for determining guard sizes, scheduling users over orthogonal resources, mitigating MUI, and configuring receiver-side equalization. For delay-based waveforms, this requirement becomes especially important because the delay-domain channel support directly affects the occupied resource region after propagation. Therefore, inaccurate or incomplete channel information may lead either to excessive guard allocation, reducing spectral efficiency, or to insufficient guard allocation, causing inter-user leakage.
}

Extensive research has been conducted on \ac{MA} techniques that enhance \ac{SE} while effectively mitigating \ac{MUI}. Among these, guard-based contiguous orthogonal MA remains a compelling solution due to its simplicity of implementation and low overhead \cite{Rakib_hadani_patent}. This approach utilizes a guard interval in the delay domain, set larger than the maximum delay spread, to minimize interference among users.
For OCDM, \ac{MU} schemes have been investigated in \cite{hossain2024design,ahmad2024dual,savaux2023ocdm}. A CP-less MU-OCDM transceiver designed for \ac{mmWave} communication and enhanced security was introduced in \cite{hossain2024design}. In \cite{ahmad2024dual}, a dual-mode MU transceiver integrating both OFDM and OCDM was studied, whereas \cite{savaux2023ocdm} examined the incorporation of multi-numerology concepts into OCDM-based systems.  
On the other hand, only a limited number of studies have addressed MA for OTSM. The work in \cite{reddy2023performance} investigated OTSM-MA along with OTFS-based \ac{SC} waveforms, highlighting the need for further research to meet the envisioned performance requirements. 
Meanwhile, \ac{MA} in OTFS systems has gained more attention. In \cite{surabhi1902multiple}, the advantages of operating in the \ac{DD} domain were leveraged to evaluate OTFS-MA performance under various resource allocation strategies in \ac{UL} scenarios for high-mobility environments, benchmarking it against OFDMA. Another OTFS-MA approach was proposed in \cite{li2020new}, where user access was performed in the spatial domain, utilizing the angle of arrival and departure as additional degrees of freedom for user separation. Further investigations on OTFS-MA techniques in the TF domain were conducted in \cite{surabhi1902multiple,khammammetti2018otfs}, where users were allocated contiguous orthogonal TF resources. The study in \cite{augustine2019interleaved} demonstrated the benefits of interleaving users in the TF domain, particularly in terms of \ac{BER} and \ac{PAPR}. Similarly, \cite{correas2023mimo} extended this concept by proposing a generalized interleaving approach to enhance sensing performance.

 However, in broadband communication systems, increased time resolution results in highly dispersive multipath channels, leading to extended delay spreads and multiple resolvable taps. Consequently, resource allocation for different UEs in delay-based waveforms necessitates the introduction of guard intervals in the delay domain to mitigate \ac{MUI}. This approach presents a trade-off: excessive guard spacing reduces \ac{SE}, while insufficient guarding leads to elevated \ac{MUI}. Moreover, the adoption of delay-based waveforms introduces challenges that were not encountered in existing systems. The problem of \ac{TM} cannot be modeled as a phase rotation in the frequency domain; rather, it results in actual shifts in resource allocation. This leads to effective channels exhibiting extended delay spreads at the BS, requiring the BS to incorporate TM effects when assigning guard intervals, which severely degrades \ac{SE}. Failure to account for it increases MUI, compromising the overall system performance.\\ This paper first introduces delay-based waveforms, including OTFS, OTSM, OCDM, and ODDM, and derives a unified input--output relation for these waveforms. We then analyze the impact of TM in uplink transmission, highlighting how the current \ac{TA} mechanisms leave some misalignment in uplink transmissions. While prior works have mainly examined timing misalignment on a per-waveform basis, this paper aims to establish a unified analytical view of how residual TM manifests across multiple delay-based waveforms in multi-user uplink. The key contributions of this work are summarized as follows:
  
 \begin{itemize}
    \item We demonstrate that in the absence of mobility, OTFS, OTSM, OCDM, and ODDM exhibit similar interactions with time-dispersive channels. Based on this observation, we derive a unified \ac{IOR}, consolidating these schemes under the umbrella of \textit{delay-based waveforms}.
    \item We discuss the \ac{TA} mechanism adopted in current standards, arguing that residual \ac{TM} among UEs persists in uplink transmission. We analyze the effect of \ac{TM} on delay-based waveforms within a unified framework, compare it with the Doppler-induced degradation observed in OFDM, and further evaluate the waveform-specific impact of joint Doppler spread and \ac{TM}.
    \item We derive analytical expressions for the interference experienced by each user and the corresponding SE, explicitly accounting for residual timing errors. Based on these insights, we propose a scheduling strategy aimed at enhancing both~\ac{SE} and \ac{SINR}. 
    \item We validate our findings through extensive simulations across various scenarios, demonstrating the accuracy of the analysis. We evaluate the average \ac{SINR} and SE using Monte Carlo simulations, showing the benefits of the proposed system. \textcolor{black}{We further extend the evaluation to waveform-specific Doppler effects, the case where the combined delay spread exceeds the CP duration, and coded BER performance.}
\end{itemize}

The remainder of this paper is organized as follows. Section~II provides the system model and explains the \ac{TA} mechanism. Section~III formulates the TM problem in \ac{MU} delay-based \ac{UL} transmissions. Section~IV presents the proposed compensation strategy for mitigating the effects of \ac{TM}, along with its theoretical foundation. Section~V provides the performance analysis, deriving expressions for \ac{SINR} and \ac{SE}. Section~VI validates the analytical results through simulations and demonstrates the gains achieved by the proposed approach. Finally, Section~VII concludes the paper with key insights and future research directions.\\
{\textit{Notation:} Bold uppercase and lowercase Latin letters denote matrices and column vectors, respectively, whereas nonbold letters denote scalars or individual entries. Bold Greek and bold calligraphic symbols also denote matrices. The operators $(\cdot)^{\mathsf T}$, $(\cdot)^\dagger$, and $(\cdot)^{-1}$ denote transpose, Hermitian transpose, and matrix inversion, respectively. The operator $\operatorname{vec}(\cdot)$ stacks the columns of a matrix into a column vector, while $\operatorname{Tr}(\cdot)$ and $\det(\cdot)$ denote the trace and determinant, respectively. The notation $[\alpha]_\beta$ denotes $\alpha$ modulo $\beta$. The Kronecker delta is denoted by $\delta(q)$, where $\delta(q)=1$ for $q=0$ and zero otherwise. The ramp function is defined as $r[t]=\max(t,0)$. The function $\operatorname{rect}(t)$ equals one for $-\frac{1}{2}\leq t<\frac{1}{2}$ and zero otherwise. The operator $\operatorname{diag}(A_1,\ldots,A_N)$ places its scalar or matrix arguments on the main (block) diagonal. The notation $\mathbb{C}^{M\times N}$ denotes the space of $M\times N$ complex-valued matrices. The symbols $\mathbf{0}_{M\times N}$ and $\mathbf{I}_N$ denote the $M\times N$ all-zero matrix and the $N\times N$ identity matrix, respectively. The symbol $\otimes$ denotes the Kronecker product, and $j$ is the imaginary unit, with $j^2=-1$.}

\begin{figure*}[t]
\centering
\includegraphics[scale = .4]{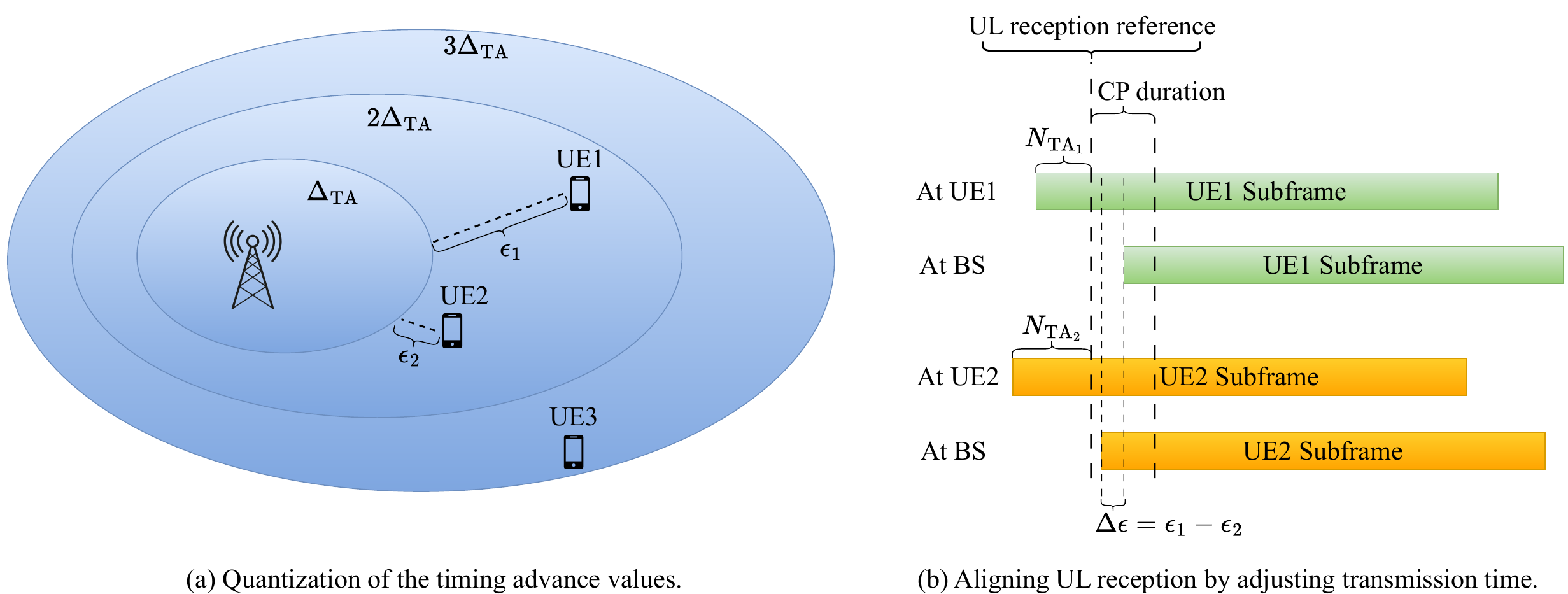}
\caption{Illustration of TA mechanism in UL systems.}
\label{fig:TA_and_how_its_quantized}
\end{figure*}
%%%%%%%%%%%%%%%%%%%%%%%%%%%%%%%%%%%%%%%%%%%%%%%%%%%%%%%%%%%%%%%%%%%%%%%%%%%%%%%%%%%%%%%%%%%%%%%%%%%%%%%%%%%%%%%%%%%%%%%%%%%%%%%
\section{System Model and Preliminaries}
\label{sec:System_model}

In this section, we first establish the unified \ac{IOR} for delay-based waveforms. Next, we examine the \ac{TA} mechanism as specified in the current 5G standard and evaluate its impact on UL transmission with these waveforms. To obtain a tractable unified analytical framework and isolate the effect of \ac{TM} in the delay domain, we first consider a Doppler-free channel model. This assumption is used only for the unified derivation; Doppler spread is later incorporated through waveform-specific effective channel representations and evaluated numerically.

\subsection{Delay-Based Waveforms}
Here, we derive a generalized \ac{IOR} framework for delay-based waveforms, including OTFS, ODDM, OCDM, and OTSM. The unification is obtained by first considering a Doppler-free channel model, under which the channel action reduces to a delay-dependent shift of the transmitted signal with path-dependent gains. This assumption allows us to isolate the impact of \ac{TM} in the delay domain and to obtain a single analytical treatment that applies across the considered waveforms. Since Doppler spread leads to waveform-specific input--output relations, its impact is treated separately later in the paper, where we extend the evaluation to include the joint impact of Doppler spread and \ac{TM} for each waveform individually. Thus, the Doppler-free assumption is used only to establish the unified analytical framework, while the simulation study captures the waveform-specific behavior under Doppler.

\subsubsection{OTFS and ODDM}
OTFS inherently multiplexes data symbols in the \ac{DD} domain. These symbols undergo a transformation via \ac{IZT} before being transmitted in the time domain. After propagating through the wireless channel, they are mapped back to the DD domain, where equalization and symbol detection are performed. Let the transmitted and received DD-domain data symbols be represented by \(\mathbf{X}^{\text{otfs}}\) and \(\mathbf{Y}^{\text{otfs}}\), respectively, with \(M\) delay bins and \(N\) Doppler bins. The \ac{IOR} for OTFS, in the absence of noise, is given by \cite{mohammed2021derivation}:
\begin{equation}
Y^{\text{otfs}}(m,n) = \textstyle \sum_{i=1}^{P} \tilde{h}_{i} X^{\text{otfs}}\left(\left[m-l_{i}\right]_{M},n\right),
\label{equ:OTFS}
\end{equation}
where the channel coefficient is expressed as \(
\Tilde{h}_i = {h}_{i}e^{\frac{j2\pi k_i (L_{\mathrm{cp}}+l-l_i)}{(M+L_{\mathrm{cp}})N}}.\)  Here, $L_{\mathrm{cp}}$ and $P$ denote the CP length and the number of channel taps, respectively; ${h}_i$ is the time domain channel coefficient. If no Doppler spread is present ($k_i = 0$), we have $\Tilde{h}_i = {h}_i$.
Rewriting \eqref{equ:OTFS} in vectorized form \cite{zegrar2023novel}, we obtain
\begin{equation}
     y^{\text{otfs}}(M\cdot n+m) = \textstyle\sum_{i=1}^{P}h_i x^{\text{otfs}}(M\cdot n+{[m-l_i]}_M),
     \label{equ:ewewew}
\end{equation}
where \(\mathbf{x}^{\text{otfs}} = \text{vec}(\mathbf{X}^{\text{otfs}})\) and \(\mathbf{y}^{\text{otfs}} = \text{vec}(\mathbf{Y}^{\text{otfs}})\). Since the received signal in an ODDM system follows the same mathematical structure as in OTFS, a similar derivation can be applied to obtain the result in \eqref{equ:ewewew}.

\subsubsection{OCDM}
The OCDM waveform utilizes chirps as basis signals for data transmission. Data symbols are multiplexed in the Fresnel domain and then transformed into the time domain via the \ac{IDFnT} before transmission. After propagation through the channel, the received signal is mapped back to the Fresnel domain, where equalization and symbol detection are performed.  
Similar to OTFS, we aim to express the IOR for OCDM in the Fresnel domain. The noise-free IOR for OCDM is given by \cite{haif2024novel}  
\begin{equation}
     y^{\text{ocdm}}(m) = \textstyle \sum_{i=1}^{P} h_{i}
     x^{\text{ocdm}}{({[m-l_{i}]}_M)},
     \label{equ:ocdm}
\end{equation}  
where \( \mathbf{x}^{\text{ocdm}} \) and \( \mathbf{y}^{\text{ocdm}} \) represent the transmitted and received signal vectors in the Fresnel domain, respectively, and \( M \) denotes the number of orthogonal chirps.

It is worth noting that even with the simplifying assumption that enabled the unification of all of these waveforms (no Doppler), the IOR for \ac{AFDM} is still different, mainly because the delays of the channels produce shifts in the affine domain that depend on the AFDM constant {$c_1$}. It is therefore interesting to see how AFDM behaves in the considered scenario, and we leave that to a future study.

\subsubsection{OTSM}
OTSM modulates data symbols in the delay-sequency domain. Let the transmitted and received symbols in this domain be represented by \(\mathbf{X}^{\text{otsm}}\) and \(\mathbf{Y}^{\text{otsm}}\), respectively, where \(M\) and \(N\) denote the number of delay and sequency bins, respectively. The IOR for OTSM, in the absence of noise, is given by \cite{otsm_transaction}
\begin{equation}
Y^{\text{otsm}}(m,n) = \textstyle \sum_{i=1}^{P} h_{i} X^{\text{otsm}}\left(\left[m-l_{i}\right]_{M},n\right).
\label{equ:otsm}
\end{equation}
Rewriting \eqref{equ:otsm} in vectorized form, we obtain
\begin{equation}
     y^{\text{otsm}}(M\cdot n+m) =\textstyle \sum_{i=1}^{P} h_i x^{\text{otsm}}(M\cdot n+{[m-l_i]}_M),
     \label{equ:otsm1D}
\end{equation}
where \(\mathbf{x}^{\text{otsm}} = \text{vec}(\mathbf{X}^{\text{otsm}})\) and \(\mathbf{y}^{\text{otsm}} = \text{vec}(\mathbf{Y}^{\text{otsm}})\).

\subsection{Generalized delay-based waveforms IOR}
A closer examination of the IOR expressions in \eqref{equ:ewewew}, \eqref{equ:ocdm}, and \eqref{equ:otsm1D} derived in the preceding subsections reveals a common structure across all waveforms. Each follows a fundamental operation of shifting and weighting the transmitted signal by the corresponding channel coefficients. This similarity arises from the fact that these waveforms are either explicitly delay-based or exhibit delay-like behavior, as in the case of OCDM. The generalized formulation can be expressed as\footnote{For OCDM, we have a single dimension represented by the variable $m$. This can easily be achieved by setting $n=0$.} 
\begin{equation}
     y(M\cdot n+m) = \textstyle\sum_{i=1}^{P} h_i x({M\cdot n+[m-l_i]}_M).
\end{equation}  
This can be expressed in matrix form as follows
\begin{equation}
     \mathbf{y} = \mathbf{H}\mathbf{x},
\end{equation}
where
\begin{equation}
    \mathbf{H} = \mathbf{I}_N\otimes\textstyle\sum_{i=1}^{P} h_{i} \boldsymbol{\Pi}_M^{l_{i}},
\end{equation}
with $\boldsymbol{\Pi}_M$ being the $M\times M$ permutation matrix and it is given~as
\begin{equation}
    \boldsymbol{\Pi}_{M}^{l_i}(m',n') = \delta([m'-n'-l_i]_{M}).
\end{equation}
The existence of the Kronecker product merely represents the vectorization of a 2-dimensional signal such as delay-Doppler. In the case of a 1-dimensional signal such as OCDM, $N$ can be set to 1, in which case the Kronecker product has no effect.

With the established generalized IOR, we can now develop a unified framework to analyze the impact of propagation delay variations in an UL system using delay-based waveforms and evaluate their effects on SE and MUI. Before proceeding with this analysis, we first provide a detailed discussion on the concept of TA employed in MU systems to mitigate the effects of varying propagation delays among different UEs.

\subsection{Timing-Advance} \label{sec:TA}

The considered scenario involves a \ac{MU} system where \ac{UE}s are distributed across different regions within a cell. Each UE's signal propagates over a unique distance, resulting in distinct propagation delays, as shown in Fig.~\ref{fig:TA_and_how_its_quantized}(a). In \ac{5G} systems, during \ac{UL} transmission, the \ac{BS} compensates for these varying propagation delays using a \ac{TA} command.  
After initial access, each UE adjusts its \ac{UL} transmission timing (either advancing or delaying it) relative to a default or estimated value. This ensures that signals from UEs with different propagation delays arrive at the \ac{BS} approximately simultaneously within the \ac{CP} duration, \(T_{\text{cp}}\), as specified in \cite{access2017requirements}. The fundamental concept of this approach is illustrated in Fig.~\ref{fig:TA_and_how_its_quantized}(b). \textcolor{black}{In this scenario, the BS commands UE1 and UE2 to transmit at $N_{\text{TA}_1}$ and $N_{\text{TA}_2}$ earlier than a given reference, respectively,  so that both signals arrive at the BS within one CP duration, with respect to the ``UL reception reference''. This does not introduce any issues for frequency-domain scheduling as long as  the residual TM values ($\epsilon_1$ and $\epsilon_2$) plus the channels' delay spreads  are kept below the CP duration.}\newline
As the locations of UEs change, the \ac{TA} values are estimated and updated accordingly. However, these updates are quantized to discrete values. If the current \ac{TA} value, \(N_{\text{TA}}\), needs to be adjusted, the \ac{BS} sends a \ac{TA} command to the UE, updating it based on an index value \(\kappa = 0,1,2,\dots,63\), according to
\begin{equation}
    N_{\text{TA}_\text{new}} = N_{\text{TA}_\text{old}} + (\kappa-31)\cdot16\cdot64\cdot \frac{T_c}{2^\mu},
    \label{eq:timeing_advance_updating}
\end{equation}
where \(\mu \) represents the numerology option, and \(T_\text{c} = 0.506~\mathrm{ns}\) is a time unit.\\
As observed in \eqref{eq:timeing_advance_updating}, the step size for updating the TA value depends on the numerology and is given by  
\begin{equation}
   \Delta_{\text{TA}} = 16\cdot64\cdot \frac{T_c}{2^\mu} \approx \frac{T_{\text{cp}}}{10}. 
\end{equation}
This quantization of TA commands results in a relatively large TM between uplink UE transmissions. For example, consider numerology \(\mu = 0\), corresponding to a \ac{SCS} of \(\Delta f = 15 \text{kHz}\). In this case, the step size, which represents the maximum allowable TM, is approximately $\Delta_{\text{TA}} \approx 0.520 ~\mathrm{\mu} s$. This is a significant value in the context of broadband transmissions, potentially impacting system performance. Moreover, practical deployment conditions can further exacerbate this TM, for example in large-cell scenarios such as \ac{NTN}. The purpose of this discussion is not to emphasize a specific numerical value, but to highlight that, in realistic multi-UE uplink operation, non-negligible residual TM among received signals is inevitable.

Fig.~\ref{fig:TA_and_how_its_quantized}(a) illustrates this idea\footnote{In the figure, we are using time and distance interchangeably as there is only a factor of the speed of light, $c$, relating them.}. TA commands can only mitigate the propagation delay differences to some extent. In this example, both UEs will have the same TA value, (i.e., $N_{\text{TA}_1}=N_{\text{TA}_2}=\Delta_{\text{TA}}$). Hence, UE1 and UE2 have residual TM, and they arrive $\Delta\epsilon=\epsilon_1-\epsilon_2$ apart at the BS. The real environment is more complicated and the TA procedure is more nuanced, but this subsection illustrated the basic idea.

%%% you can define the lengths of the in/out signals, for each waveform.
\section{Problem Description}
As mentioned above, we are interested in an \ac{UL} system involving one \ac{BS} and $K_u$ UEs, where the signals from different UEs experience different propagation delays. After ensuring that all UE transmissions arrive within the same CP (through a TA operation), there remains a residual delay with respect to the perfect alignment with the reception time of the BS. We denote these delays by $\epsilon_1,\ldots,\epsilon_{K_u}$. These residual delays can easily be fractional, leading to leakage and interference scenarios that can be a major obstacle for most of these waveforms. In order not to deviate from the focus of this work, and to enable derivations of closed-form expressions, we consider the delays to be integers. We assess the accuracy of this assumption in Section~\ref{sec:simulation}. Therefore, let the normalized residual delay due to TM for the $k$-th UE be given by
\begin{equation}
    \lambda_k = {B_w}\cdot\epsilon_k,
\end{equation} 
where $B_w$ is the total utilized bandwidth. The IOR for this UE under the assumption of different propagation delays is given~by\footnote{For vectors and matrices, the user index is written as a subscript in parentheses, e.g., $\mathbf{x}_{(k)}$ and $\mathbf{H}_{(k)}$. This avoids ambiguity with subscripts used to denote matrix sub-blocks (written without parentheses). We avoid using superscripts for user indexing to preserve superscripts for algebraic operations (e.g., $(\cdot)^{-1}$, $(\cdot)^{\dagger}$). For scalars, we use the conventional subscript $h_k$ since no block-index conflict arises and parentheses would unnecessarily clutter the expressions.} 
\begin{equation}
     \mathbf{y}_{(k)}= \mathbf{H}_{(k)}\mathbf{x}_{(k)},
     \label{eq:single_in_out}
\end{equation}
where 
\begin{equation}
    \mathbf{H}_{(k)} = \mathbf{I}_N\otimes\sum_{i=1}^{P_k} h_{k,i} \boldsymbol{\Pi}_M^{l_{k,i}+\lambda_k},
    \label{eq:channel_mat_lambda}
\end{equation}
and the received signal at the BS is the sum over all $K_u$ UEs, given by
\begin{equation}
     \mathbf{y} = \sum_{k=1}^{K_u}\mathbf{H}_{(k)}\mathbf{x}_{(k)}.
     \label{eq:general_in_out}
\end{equation}
Here, $\mathbf{x}_{(k)}$ is the transmitted symbols mapped to the allocated resources of the $k$-th UE. For the channel coefficients and delays, the same notation as in Section~\ref{sec:System_model} is used, with the inclusion of the user index. \\
The additional $\lambda_k$ in \eqref{eq:channel_mat_lambda} gives a different perspective on resource allocation in the considered waveforms. The resulting effective channels are no longer affected only through an additional linear phase; rather, TM produces actual shifts in the allocated resources. We now mention briefly two types of issues that can arise based on these nonzero $\lambda_k$ values.
\subsection{Effect of Misconfigured Guards on 
MUI}\label{subse:MUI_problem}

In this setup, assume the \ac{BS} has knowledge of the actual \ac{DL} channel of each \ac{UE}, which has been estimated by the \ac{UE} and reported back to the \ac{BS}. In this case, a guard interval \( g_k \) along delay is selected based on the \ac{AMDS} \( l_{k,\max} \) of the corresponding \ac{UE}. These channels do not incorporate the \ac{TM} error \( \lambda_k \) and assume that all \acp{UE} are ideally positioned along the quantized circles illustrated in Fig.~\ref{fig:TA_and_how_its_quantized}(a). This results in \ac{MUI}, where each \ac{UE} experiences interference leakage proportional to \( \lambda_k \) from the remaining \acp{UE}.\\
The core challenge becomes mitigating this MUI. A straightforward solution is to assign larger guard intervals \( g_k \) to absorb the non-zero timing misalignment \( \lambda_k \). However, as elaborated in the following subsection, this approach is spectrally inefficient, as it substitutes interference mitigation with bandwidth overhead, thus trading off spectral efficiency for robustness.

\subsection{Effect of Misconfigured Guards on SE}\label{subse:SE_problem}

An alternative, and arguably more practical, approach assumes that the \ac{BS} relies solely on \ac{UL} channel estimates for each \ac{UE}. Under this assumption, the guard assigned to the \(k\)-th \ac{UE} is set to \(l_{k,\max} + \lambda_k\), corresponding to the \ac{MMDS} for that user. This design ensures that the \ac{BS} remains protected from interference even in the presence of a non-zero \(\lambda_k\). However, the applied guard is larger than what would have been needed in the ideal case where \(\lambda_k = 0\). As a result, the system suffers a loss in \ac{SE}, potentially leading to degraded performance bounds. \\
The resulting design challenge is to maximize system \ac{SE} while ensuring minimal interference, given the practical constraint of non-zero \(\lambda_k\) values.

\begin{figure*}[t]
\centering
\includegraphics[scale = .26]{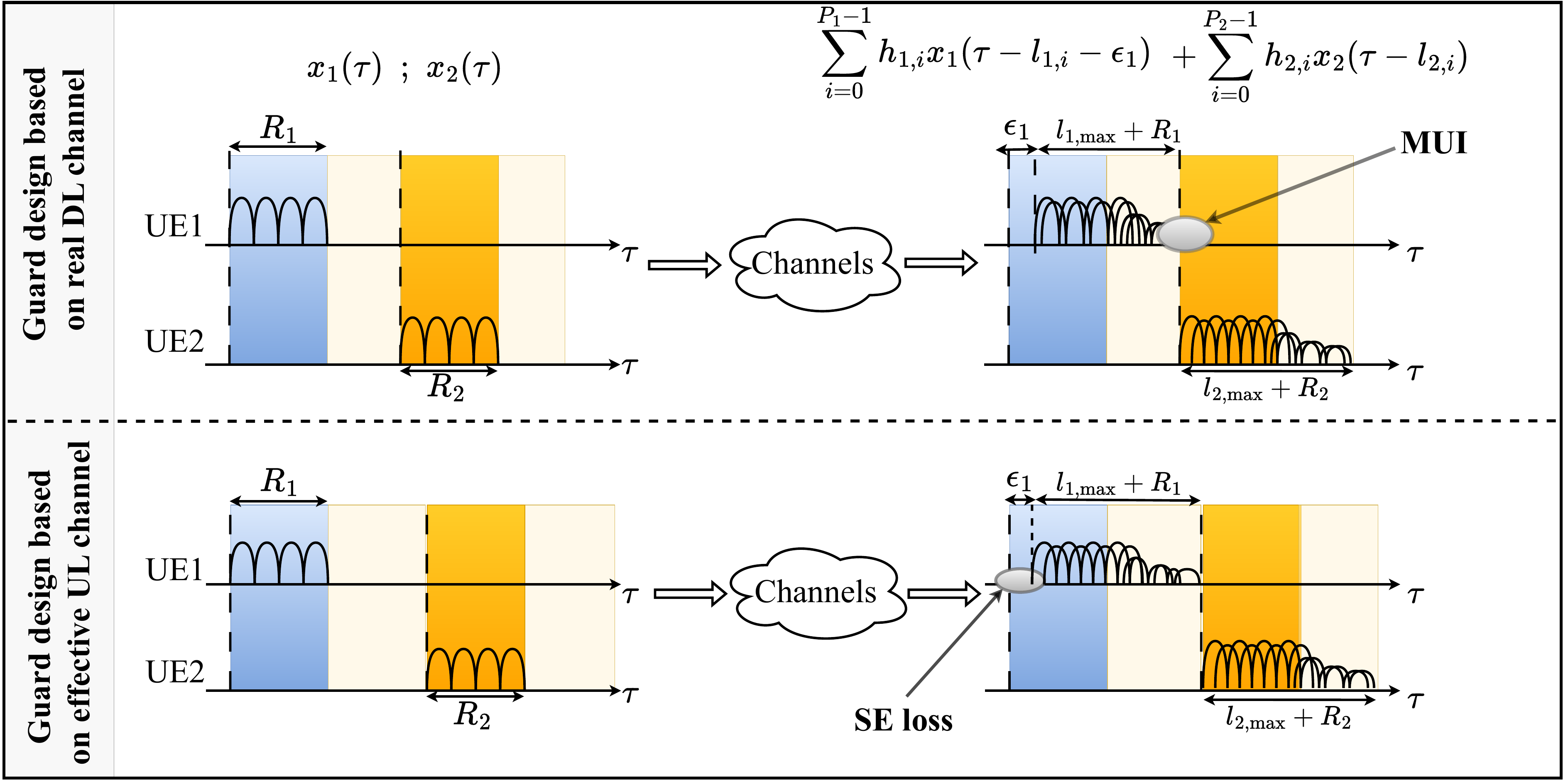}
\caption{Comparison of TM effect on OFDM and delay-based waveforms.}
\label{fig:The_effect_of_TM}
\end{figure*}

Fig.~\ref{fig:The_effect_of_TM} shows the occupied resources  for two UEs: UE1 with a TM of $\epsilon_1$ and UE2 with no TM. Guards are chosen based on the two alternatives described above. The effect is clearly shown, as UE1 causes leakage over the delay resources of UE2, thereby degrading SINR for both users, for the first alternative. For the other alternative, the guard is chosen to combat both channel delay spread and TM. As can be seen in the received UL signal, some resources remain unutilized by any UE, degrading the sum SE of the overall system. 

\subsection{Benchmark: OFDM Under TM}
\label{sec:including_doppler_for_ofdm}

\textcolor{black}{The focus of this paper is to analyze the effect of \ac{TM} within a unified framework that applies to the considered delay-based waveforms. To enable this unified \ac{IOR}, the main analysis neglects Doppler spread, since including Doppler leads to waveform-specific input--output relations. OFDM, however, is treated separately as a benchmark, which allows Doppler spread to be incorporated explicitly. This is also necessary because, while OFDM is robust to within-CP TM, Doppler spread is a dominant source of interference for OFDM and therefore provides a meaningful reference impairment. Accordingly, for the OFDM baseline, we include Doppler-induced time selectivity. Later, in Section~\ref{sec:doppler_for_each_waveform}, Doppler spread is also incorporated for each delay-based waveform through its corresponding waveform-specific effective channel representation. To this end, we define the time-delay channel of the \(u\)-th UE as}\\

\begin{equation}
    \mathbf{H}_{(u)}^\text{TD} = \text{diag}\left (\mathbf{H}_{(u);0}^\text{TD},...,\mathbf{H}_{(u);N-1}^\text{TD}  \right),
    \label{equ:Ht}
\end{equation}
where the $n$-th block is given by
\begin{equation}
    \mathbf{H}_{(u);n}^\text{TD} =\sum_{i=1}^{P_u} h_{u,i} \boldsymbol{\Pi}^{l_{u,i}+\lambda_u}_{M}\boldsymbol{\Omega}^{k_{u,i},n}_{M},
\end{equation}
where $\boldsymbol{\Omega}^{k_{u,i},n}_{M}(m',n') = \boldsymbol{\Omega}^{k_{u,i}}_{M}(m',n') ~\text{for}~m'\in [n(M+L_{\text{cp}})+L_{\text{cp}},(n+1)(M+L_{\text{cp}})-1]$.
Then, the \ac{TF} effective channel (representing the IOR for OFDM) can be obtained by
\begin{equation}
   \mathbf{H}_{(u)} =(\mathbf{I}_N\otimes \mathbf{F}_M) \mathbf{H}_{(u)}^\text{TD}( \mathbf{I}_N \otimes \mathbf{F}^\dagger_M).
   \label{equ:Htf}
\end{equation}

This is also block-diagonal, with each block given by 

\begin{equation}
\begin{aligned}
\mathbf{F}_M\mathbf{H}_{(u);n}^\text{TD} \mathbf{F}^\dagger_M&=\mathbf{F}_M\sum_{i=1}^{P_u} h_{u,i} \boldsymbol{\Pi}^{l_{u,i}+\lambda_u}_{M}\boldsymbol{\Omega}^{k_{u,i},n}_{M}\mathbf{F}^\dagger_M\\
&=\mathbf{F}_M\boldsymbol{\Pi}^{\lambda_u}_{M}\mathbf{F}^\dagger_M\mathbf{F}_M\sum_{i=1}^{P_u} h_{u,i} \boldsymbol{\Pi}^{l_{u,i}}_{M}\boldsymbol{\Omega}^{k_{u,i},n}_{M}\mathbf{F}^\dagger_M.\\
\end{aligned}
\end{equation}

In the absence of Doppler, we would have a diagonal matrix, $\mathbf{F}_M\boldsymbol{\Pi}^{\lambda_u}_{M}\mathbf{F}^\dagger_M\mathbf{F}_M\sum_{i=1}^{P_u} h_{u,i} \boldsymbol{\Pi}^{l_{u,i}}_{M}\mathbf{F}^\dagger_M$, which indicates that no leakage occurs across frequency-domain resources. In this case, the linear phase introduced by $\mathbf{F}_M\boldsymbol{\Pi}^{\lambda_u}_{M}\mathbf{F}^\dagger_M$ only modifies the diagonal entries through an element-wise phase rotation and does not destroy orthogonality. However, the presence of Doppler violates this idealized diagonal structure, making the system suboptimal.

\section{Proposed Resource Allocation Scheme for Delay-Based Waveforms Under TM}\label{sec:Proposed}

In the system model under consideration, \acp{UE} are allocated orthogonal resource regions in a contiguous manner along the delay domain, with optional guard intervals between users\footnote{The unified analytical development focuses on users allocated along the delay domain. Other domain-based access strategies, such as Doppler-domain allocation for OTFS, may also be considered depending on the channel conditions and scheduling objectives. A general multi-domain scheduling framework is beyond the scope of this work; here, we focus on delay-domain allocation because it directly exposes the impact of residual \ac{TM}.}. We adopt a \ac{DeDMA} framework, where user separation occurs along the delay axis. In this framework, the \(k\)-th \ac{UE} is assigned the resource set \( \{[T_k]_M, [T_k+1]_M, \dots, [T_k + \mathcal{R}_k - 1]_M\} \), where \(\mathcal{R}_k\) denotes the number of delay bins allocated to the \(k\)-th user.
Subsequently, the \((k+1)\)-th \ac{UE} is allocated the next contiguous set of resources given by \(\{ [T_{k+1}]_M, \dots, [T_{k+1} + \mathcal{R}_{k+1} - 1]_M \}\), where \( T_{k+1} = T_k + \mathcal{R}_k + g_k \). This ensures that users are placed sequentially along the delay axis, separated by guard intervals \( g_k \) to mitigate inter-user interference.\\
We define \( M'_k = \mathcal{R}_k + g_k \) as the total delay-domain resource footprint per \ac{UE}, including its associated guard. For simplicity, we assume uniform allocation across users, i.e., \( M'_k = M' \) for all \(k\). Then, the number of users that can be supported within the total available delay dimension \(M\) is given by \( \beta = M / M' \). \\
The expression for the allocation matrix is given as 
\begin{equation}
     \mathbf{Q}_{(k)}= \mathbf{I}_{N}\otimes(\mathbf{e}_{k}\otimes \mathbf{I}_{M'}),
     \label{eq:allocation_matrix}
\end{equation}
where \( \mathbf{e}_k\) denotes the \(k\)-th standard basis vector in \( \mathbb{R}^\beta \) defined by its $i$-th element as $\mathbf{e}_k(i) = \delta(i-k)$.
This matrix $\mathbf{Q}_{(k)}$ serves as a mathematical representation of the way resources are allocated to the $k$-th UE, such that \begin{equation}
    \mathbf{x}_{(k)} = \mathbf{Q}_{(k)}\mathbf{G}_{(k)}\mathbf{u}_{(k)},
\end{equation}
where $\mathbf{u}_{(k)}$ represents the symbol vector, $\mathbf{x}_{(k)}$ represents symbols in $\mathbf{u}_{(k)}$ mapped to their corresponding resources, and $\mathbf{G}_{(k)} \in \mathbb{C}^{N(\mathcal{R}_k+ g_k)\times N\mathcal{R}_k}$ represents the guard allocation along the delay and it is defined by
\begin{equation}
     \mathbf{G}_{(k)}= \mathbf{I}_{N}\otimes{[\mathbf{I}_{ \mathcal{R}_k} ,\mathbf{0}_{\mathcal{R}_k \times g_k}]}^\dagger.
     \label{eq:guard_matrix}
\end{equation}

It is possible to estimate \ac{TM} for each UE, including fractional offsets, using existing synchronization and delay-estimation techniques. In this work, we do not benchmark these estimators in terms of accuracy or complexity. Instead, as will be shown in Section~\ref{sec:simulation}, we directly study the impact of residual timing errors that remain after such estimation and compensation procedures. A dedicated evaluation of estimator performance (e.g., bias/variance–complexity trade-offs) is outside our scope, and we refer interested readers to the existing literature.

One representative approach is the linear-phase estimation method in \cite{kay2002fast}, which was applied to estimate residual timing offsets in OFDM systems in \cite{hsieh1998channel}. This principle readily extends to other multicarrier and delay-based waveforms when frequency-domain pilots (or pilot-equivalent observations) are available. For example, consider an OTFS frame that includes impulse pilots in the \ac{DD} domain. Assuming the pilots of different users are orthogonally multiplexed (hence not subject to MUI during the estimation stage), the receiver can obtain a pilot-bearing representation in the frequency domain after the usual processing (e.g., Doppler estimation/compensation followed by an appropriate transform). In that domain, a residual timing offset manifests predominantly as a linear phase slope across frequency, enabling the use of linear-phase estimators to recover the user-specific timing misalignment \cite{hsieh1998channel}. Since TM is an UL phenomenon arising from multi-UE propagation and scheduling, such estimation is naturally performed at the BS. Depending on the assumed temporal dynamics of the propagation delays and residual offsets, the resulting estimates can be fed back to the UEs for pre-compensation prior to transmission, thereby limiting residual misalignment and improving \ac{SINR} and \ac{SE}.

Alternatively, compensation can be performed entirely at the BS without feedback. In this case, after estimating the TM values, the received uplink signal can be modeled as an equivalent single-user signal corrupted by deterministic timing-induced coupling terms, similar in spirit to \cite{reddy2023performance}. Under this viewpoint, what appears as inter-user interference can be reinterpreted as structured \ac{ISI}/\ac{ICI} that is amenable to equalization (e.g., within an OTFS receiver). The main drawback is complexity: the effective input–output relation expands with the number of users, increasing the dimension of the corresponding \ac{IOR}. In the present setting, however, the TM is sufficiently small that only adjacent allocations contribute significant leakage; hence, it is often adequate to restrict the model to the two nearest neighboring users, potentially making the receiver-side compensation computationally tolerable.

The issues outlined in Subsections~\ref{subse:MUI_problem} and~\ref{subse:SE_problem} both stem from a common and practical limitation: the inherently asynchronous nature of \ac{UL} transmissions. Such timing mismatches are inevitable in real-world deployments. However, it is important to emphasize that these problems are not rooted in the waveform structure or the underlying \ac{MA} techniques themselves. Rather than modifying the waveform or access mechanism, we propose a lightweight solution that introduces a pre-transmission rescheduling step at the \ac{UE} side.

The proposed approach retains the original resource allocation structure but introduces a timing compensation mechanism at the \ac{UE}. This mechanism applies a pre-transmission delay adjustment to counteract the residual timing offset \( \lambda_k \), thereby mitigating both \ac{SE} degradation and \ac{MUI} simultaneously, as described in Section~\ref{sec:Proposed}.
In particular, each \(k\)-th \ac{UE} leverages residual timing offset estimation \( \tilde{\lambda}_k \), fed back by the BS as described above, to apply a correction prior to transmission. The objective is to minimize the effective timing mismatch \( \delta_k = \lambda_k-\tilde{\lambda}_k \), ideally achieving \( \delta_k = 0 \). We assume that each \ac{UE} obtains its timing offset within a bounded accuracy of \( \mathcal{T} \) samples, i.e., \( |\delta_k| \leq \mathcal{T} \) for all \(k = 1, \dots, K_u \).

Accordingly, instead of transmitting the original signal \( \mathbf{x}_{(k)} \), the UE transmits a rescheduled signal \( \tilde{\mathbf{x}}_{(k)} \), computed as
\begin{equation}
     \tilde{\mathbf{x}}_{(k)} =  (\mathbf{I}_N \otimes \mathbf{\Pi}_M^{-\tilde{\lambda}_k}) \mathbf{x}_{(k)} = (\mathbf{I}_N \otimes \mathbf{\Pi}_M^{-\tilde{\lambda}_k}) \mathbf{Q}_{(k)} \mathbf{G}_{(k)} \mathbf{u}_{(k)}.
     \label{eq:proposed_mechanism}
\end{equation}
The total received signal at the \ac{BS} from all \acp{UE} is then expressed as
\begin{equation}
     \mathbf{y} = \sum_{k=1}^{K_u} \mathbf{H}_{(k)} (\mathbf{I}_N \otimes \mathbf{\Pi}_M^{-\tilde{\lambda}_k}) \mathbf{x}_{(k)}.
     \label{eq:rx_BS_signal_acc_to_proposed}
\end{equation}
It follows from \eqref{eq:rx_BS_signal_acc_to_proposed} that the effective channel observed for the \(k\)-th \ac{UE} becomes:
\begin{equation}
     \mathbf{H}'_{(k)} = \mathbf{H}_{(k)} ( \mathbf{I}_N \otimes \mathbf{\Pi}_M^{-\tilde{\lambda}_k}) = \mathbf{I}_N \otimes \sum_{i=1}^{P_k} h_{k,i} \mathbf{\Pi}_M^{l_{k,i} + \lambda_k - \tilde{\lambda}_k}.
     \label{eq:proposed_general_channel}
\end{equation}
In the subsequent analysis, all derived expressions apply to both the conventional and proposed schemes. The distinction lies in the residual timing offset: for the conventional case, \( \delta_k = \lambda_k \), while in the proposed scheme, it is reduced to \( \delta_k = \lambda_k - \tilde{\lambda}_k \).

\section{Performance Analysis} \label{sec:analysis}

To obtain expressions for the MUI and the \ac{SE}, it is helpful to first establish some lemmas\footnote{The lemmas in this section are primarily technical tools; readers mainly interested in the results may skim them and refer back when needed.}. \\
\noindent \textbf{Lemma 1:} Given the matrix $\bm{\mathcal{X}}_{(k)}$, defined by 
\begin{equation}
     \bm{\mathcal{X}}_{(k)} = \begin{bmatrix}
\mathbf{I}_{\mathcal{R}_k\times \mathcal{R}_k}& \mathbf{0}_{\mathcal{R}_k\times g_k}  \\
\mathbf{0}_{g_k \times \mathcal{R}_k}& \mathbf{0}_{g_k\times g_k}\\ 
\end{bmatrix} ,
\end{equation}
 we have the following equality
\begin{equation}
     \mathbf{G}_{(k)}\mathbf{G}^\dagger_{(k)} = \mathbf{I}_{N}\otimes \bm{\mathcal{X}}_{(k)},  
\end{equation}
\noindent \textbf{Proof:}
The proof is found by applying the Kronecker-of-product property using the definition in \eqref{eq:guard_matrix}.\hfill$\blacksquare$

\noindent \textbf{Lemma 2:} Let $\mathbf{A}_{(k)}  = \mathbf{Q}_{(k)}\mathbf{Q}^\dagger_{(k)}$, and define the matrix $\mathbf{E}_k =\mathbf{e}_k{(\mathbf{e}_k)}^\dagger \in \mathbb{C}^{\beta \times \beta}$. This matrix has only one non-zero element, which is the $k$-th element on the diagonal, and it is equal to 1. Then, we have
\begin{equation}
     \mathbf{A}_{(k)}  = \mathbf{I}_{N}\otimes \underbrace{(\mathbf{E}_k \otimes \mathbf{I}_{M'})}_{\bm{\mathcal{A}}_{(k)}}   ,
\end{equation}
where $\bm{\mathcal{A}}_{(k)}$ is a (block) diagonal matrix, with all blocks $\bm{\mathcal{A}}_{(k)}$ being identical with the elements satisfying 
\begin{equation}
     \bm{\mathcal{A}}_{(k)}(i,j)  = \delta(i-j)\text{rect}\left(\frac{i-M'/2-(k-1)M'}{M'}\right) .
\end{equation}

\noindent \textbf{Proof:} The proof is found directly using the definition in \eqref{eq:allocation_matrix}. \hfill$\blacksquare$\\

In a similar manner, let $\mathbf{{B}}_{(k)}  = \mathbf{Q}_{(k)}\mathbf{G}_{(k)}\mathbf{G}^\dagger_{(k)}\mathbf{Q}^\dagger_{(k)}$, then we can show the following
\begin{equation}
     \mathbf{B}_{(k)}  = \mathbf{I}_{N}\otimes  \underbrace{(\mathbf{E}_k \otimes \bm{\mathcal{X}}_{(k)})}_{\bm{\mathcal{B}}_{(k)}}.
\end{equation}
The matrix $\bm{\mathcal{B}}_{(k)}$ is (block) diagonal with the $N$ blocks $\bm{\mathcal{B}}_{(k)}$ satisfying
\begin{equation}
     \bm{\mathcal{B}}_{(k)}(i,j)  = \delta(i-j)\text{rect}\left(\frac{i-\mathcal{R}_k/2-(k-1)M'}{\mathcal{R}_k}\right) .
     \label{eq:cor}
\end{equation}
Let ${\bm{\mathcal{B}}}'_{(k)}(i,j) = \boldsymbol{\Pi}^{-l'}_M\bm{\mathcal{B}}_{(k)}(i,j)$ and ${\bm{\mathcal{A}}}'_{(k)}(i,j) = \bm{\mathcal{A}}_{(k)}(i,j)\boldsymbol{\Pi}^{-l}_M$, by applying the circular shifting to the columns and rows of $\bm{\mathcal{B}}_{(k)}$  and $\bm{\mathcal{A}}_{(k)}$, it is clear that
\begin{equation}
     {\bm{\mathcal{B}}}'_{(k)}(i,j)  = \delta({[i+l']}_M-j)\text{rect}\left(\frac{j-\mathcal{R}_k/2-(k-1)M'}{\mathcal{R}_k}\right) ,
\label{eq:cor1}
\end{equation}
and
\begin{equation}
     {\bm{\mathcal{A}}}'_{(k)}(i,j)  = \delta({[i+l]}_M-j)\text{rect}\left(\frac{j-\mathcal{R}_k/2-(k-1)M'}{M'}\right) .
     \label{eq:cor2}
\end{equation}
\noindent \textbf{Lemma 3:}
With the above setup, we can show the following
\begin{equation}
\begin{aligned}
    \text{Tr}({\bm{\mathcal{B}}_{(k-1)}}^{\dagger}{\bm{\mathcal{A}}_{(k)}}) 
    =r[l-g_{k-1}]\delta(l-l'),
\end{aligned}
\end{equation}
which is only non-zero for $l=l'$.

\noindent \textbf{Proof:} The proof is given in Appendix~A.

\noindent \textbf{Lemma 4:} The trace of $\bm{\mathcal{B}}'_{(k)}$ can be found as
\begin{equation}
\begin{aligned}
    \text{Tr}({\bm{\mathcal{B}}'_{(k)}}^{\dagger}{\bm{\mathcal{A}}'_{(k)}}) 
    =r\big[M'-\big(l+r[g_k-l]\big)\big]\delta(l-l'),
\end{aligned}
\end{equation}
which is nonzero only for $l=l'$.\\
\noindent \textbf{Proof:} The proof is given in Appendix~B.

Equipped with these lemmas, we can find expressions for \ac{SINR} and \ac{SE} in terms of $\delta_k$, the residual misalignment.

\subsection{MUI for Delay-Based Waveforms}
\label{sec:MUI_delay_waveforms}
Here, we treat the delay-based waveforms according to the unified \ac{IOR} we derived. We begin by establishing two foundational theorems that enable the derivation of the SINR expression for the \(k\)-th \ac{UE}, providing analytical insight into the system's interference characteristics. The noise-free received signal component associated with the \(k\)-th \ac{UE} can be expressed as:  
\begin{equation}
\begin{aligned}
    \mathbf{v}_{(k)} &= \mathbf{Q}_{(k)}^\dagger \sum_{j=1}^{K_u} \mathbf{H}_{(j)} \mathbf{x}_{(j)} \\
    &= \underbrace{\mathbf{Q}_{(k)}^\dagger \mathbf{H}_{(k)} \mathbf{x}_{(k)}}_{\text{desired signal}} + \underbrace{\mathbf{Q}_{(k)}^\dagger \sum_{\substack{j=1 \\ j \neq k}}^{K_u} \mathbf{H}_{(j)} \mathbf{x}_{(j)}}_{\text{MUI}},
\end{aligned}
\label{eq:UL-rx_K_uE}
\end{equation}  
where the first term denotes the desired signal component and the second term captures the multi-user interference (MUI) impacting the \(k\)-th \ac{UE}.

To characterize the interference leakage from the \((k-1)\)-th \ac{UE}, we define the matrix \( \mathbf{\Xi}_{(k)} = \mathbf{Q}_{(k)}^\dagger \mathbf{H}_{(k-1)} \mathbf{Q}_{(k-1)} \mathbf{G}_{(k-1)} \). Let \( l'_{k-1,i} = l_{k-1,i} + \delta_{k-1} \), accounting for the delay and residual timing mismatch of the interfering user. For non-zero \( \delta_{k-1} \), there exists non-negligible interference from the \((k-1)\)-th user to the \(k\)-th user\footnote{We assume \( M' \) is sufficiently large to ensure that leakage is limited to adjacent users only. This constraint is critical to maintaining practical system feasibility.}. The resulting interference power is quantified by the  Frobenius norm of \( \mathbf{\Xi}_{(k)} \), and is given by
\begin{equation}
    \zeta_k = \frac{1}{K'} \text{Tr}({(\mathbf{\Xi}_{(k)})}^\dagger \mathbf{\Xi}_{(k)}),
    \label{equ:hdhdh}
\end{equation}
where $K'=M'N$ and \( \zeta_k \) serves as a metric for the inter-user interference imposed on the \(k\)-th \ac{UE} due to imperfect time alignment of the \((k-1)\)-th \ac{UE}.

\noindent \textbf{Theorem 1:}
With the general channel matrix given by \eqref{eq:proposed_general_channel}, the interference experienced by the $k$-th UE $\zeta_k$ can be expressed as\footnote{$r[t]$ is the ramp function defined in the notation.} 
\begin{equation}
    \zeta_k=\frac{N}{K'} \sum_{i = 1}^{P_{k-1}}r\bigg[l_{k-1,i}+\delta_{k-1}-g_{k-1}\bigg]{|h_{k-1,i}|}^2.
    \label{eq:interferece_expression_prop}
\end{equation} 
\noindent \textbf{Proof:} The proof is given in Appendix~C. 

We now shift our focus to analyzing the received signal strength \( \sigma_k \) for the \(k\)-th \ac{UE}. Define the matrix \( \mathbf{\Gamma}_{(k)} = \mathbf{Q}_{(k)}^\dagger \mathbf{H}_{(k)} \mathbf{Q}_{(k)} \mathbf{G}_{(k)} \), then, analogous to the interference expression, the signal strength can be computed using the squared Frobenius norm as  
\begin{equation}
    \sigma_k = \frac{1}{K'} \text{Tr}( \mathbf{\Gamma}_{(k)}^\dagger \mathbf{\Gamma}_{(k)} ).
\end{equation}

\noindent \textbf{Theorem 2:}  
Given the general channel model in \eqref{eq:proposed_general_channel}, the signal strength \( \sigma_k \) of the \(k\)-th \ac{UE} is given by
\begin{equation}
    \sigma_k = \frac{N}{K'} \sum_{i=1}^{P_k} r\big[M' - l_{k,i} - \delta_k - r[g_k - l_{k,i} - \delta_k]\big] |h_{k,i}|^2,
\label{eq:ss_expression_prop}
\end{equation}
where \( r[\cdot] \) denotes the ramp function.

\noindent \textbf{Proof:}  
The proof follows the same steps as in \textbf{Theorem 1}, utilizing \textbf{Lemma 4} instead of \textbf{Lemma 3}. \hfill\(\blacksquare\)

\textcolor{black}{Theorems 1 and 2 provide direct design insights for delay-domain resource allocation under residual timing misalignment. From Theorem 1, the interference from the $(k-1)$-th UE to the $k$-th UE is governed by the excess effective delay
$l_{k-1,i}+\delta_{k-1}-g_{k-1}$. If this quantity is non-positive for all channel paths, the ramp function becomes zero and no leakage from the neighboring UE is observed. Otherwise, the interference increases with the part of the effective channel support that exceeds the allocated guard interval. Hence, residual timing misalignment acts as an additional delay-domain shift that enlarges the effective support of the interfering UE and may create MUI even when the physical channel delay spread alone is properly guarded.\\
Theorem 2 gives the corresponding implication for the desired signal component. When the effective delay spread of the desired UE is contained within the protected region, the useful signal contribution over the allocated data resources is preserved. However, when $l_{k,i}+\delta_k$ exceeds the available guard, part of the desired signal is shifted outside the intended allocation, which reduces the useful received signal strength. Therefore, residual timing misalignment has a twofold effect: it increases the interference leakage from adjacent users and can also reduce the desired signal power of the affected UE.\\
These observations motivate the proposed compensation strategy. A straightforward solution is to increase the guard interval such that it covers both the channel delay spread and the residual timing offset. However, this reduces the number of useful delay-domain resources and degrades spectral efficiency. Instead, the proposed scheme aims to reduce the residual timing error $\delta_k$ through pre-transmission timing compensation. By reducing $\delta_k$, the effective delay support is compressed back toward the intended allocation region, thereby reducing MUI without requiring excessive guard overhead.}

With the expressions from \textbf{Theorem 1} and \textbf{Theorem 2}, the SINR for the \(k\)-th \ac{UE} can be expressed as \cite{tse2005fundamentals}:
\begin{equation}
\begin{aligned}
    \mathrm{SINR}_k = \frac{\sigma_k}{\gamma^{-1} + \zeta_k},
\end{aligned}
\label{eq:SINR_prop}
\end{equation}
where \(\zeta_k\) quantifies the interference, $\gamma = E_s/N_0$ denotes the \ac{SNR} and \(\sigma_k\) reflects the signal power.

\subsection{MUI for OFDM Waveform}
To maintain consistency with the delay-domain waveform analysis, we adopt an analogous allocation structure in the frequency domain, where the frequency axis replaces the delay axis (i.e., a \ac{FDMA} scheme). Optional guards can be used to mitigate interference caused by Doppler spread. As previously discussed, for OFDM, we include the Doppler spread in the analysis.

We first compute the MUI experienced by the \(k\)-th \ac{UE}. Then, we derive an expression for the received signal strength, and subsequently, the SINR for the \(k\)-th \ac{UE}.

To begin with, we decompose the frequency-domain channel matrix of the \(j\)-th \ac{UE} into its block-diagonal components. From \eqref{equ:Htf}, we have

\begin{equation}
\begin{aligned}
    \mathbf{H}_{(j)} = \sum_{n=0}^{N-1}\bm{\mathcal{E}}_n\otimes\hat{\mathbf{H}}_{(j);n},
        \label{equ:TF koko}
\end{aligned}
\end{equation} 
where $\bm{\mathcal{E}}_n \in \mathbb{C}^{N \times N}$ has only one nonzero element, at the $n$-th diagonal entry, which is equal to $1$.\\
In general, frequency domain guards can be applied. Using the mixed-product property,  we have
\begin{equation}
\mathbf{G}^\dagger_{(k)}{(\mathbf{Q}_{(k)})}^\dagger\mathbf{H}_{(j)}\mathbf{Q}_{(j)}\mathbf{G}_{(j)} ={(\mathbf{Q}'_{(k)})}^\dagger\mathbf{H}_{(j)}\mathbf{Q}'_{(j)}=
    \sum_{n=0}^{N-1}{\bm{\mathcal{E}}}_{n} \otimes  {\mathbf{H}}^\prime_{n},
    \label{equ:summ}
\end{equation}
where $\mathbf{Q}'_{(j)}=\mathbf{Q}_{(j)}\mathbf{G}_{(j)}$, ${\mathbf{H}}^\prime_{n} =\hat{\mathbf{G}}^\dagger_{(k)}\hat{\mathbf{H}}_{(j);n}\hat{\mathbf{G}}_{(j)}$, and $\hat{\mathbf{G}}_{(k)}=\bigg({\mathbf{e}_{k}}\otimes\mathbf{I}_{M^\prime}\bigg){[\mathbf{I}_{ \mathcal{R}_k} ,\mathbf{0}_{\mathcal{R}_k \times g_k}]}^\dagger$.\\
At the \ac{BS}, the interference power \cite{tse2005fundamentals} observed over the $k$-th UE’s resources from the $j$-th UE is given by the squared Frobenius norm of $ {(\mathbf{Q}'_{(k)})}^\dagger\mathbf{H}_{(j)}\mathbf{Q}'_{(j)} $, as 
\begin{equation}
    \text{T}_{k,j} \overset{\triangle}{=} \text{Tr} \big[ {\big( {(\mathbf{Q}'_{(k)})}^\dagger\mathbf{H}_{(j)}\mathbf{Q}'_{(j)}\big)}^\dagger \big( {(\mathbf{Q}'_{(k)})}^\dagger\mathbf{H}_{(j)}\mathbf{Q}'_{(j)}\big) \big].
\end{equation} 

We proceed by expanding the expression as follows:
\begin{equation}
    \begin{aligned}
        \text{T}_{k,j} &= \text{Tr}\bigg( \sum_{n=0}^{N-1}{\bm{\mathcal{E}}}_{n} \otimes  {{\mathbf{H}}^\prime_{n}}^\dagger \sum_{n=0}^{N-1}{\bm{\mathcal{E}}}_{n} \otimes  {\mathbf{H}}^\prime_{n}\bigg)\\
        &\overset{a}{=}\text{Tr}\bigg(\sum_{n=0}^{N-1}\sum_{m=0}^{N-1}{\bm{\mathcal{E}}}_m{\bm{\mathcal{E}}}_n \otimes  {({\mathbf{H}_m')}^\dagger}{\mathbf{H}_n'}\bigg)\\
        &\overset{b}{=}\text{Tr}\bigg(\sum_{n=0}^{N-1}{\bm{\mathcal{E}}}_n \otimes  {({\mathbf{H}_n')}^\dagger}{\mathbf{H}_n'}\bigg)\\
        &\overset{c}{=}\sum_{n=0}^{N-1}\text{Tr}\left({({\mathbf{H}_n')}^\dagger}{\mathbf{H}_n'}\right)\\
        &\overset{d}{=}\sum_{n=0}^{N-1}\text{Tr}\Bigg(\hat{\mathbf{G}}_{(j)}{(\hat{\mathbf{G}}_{(j)})}^\dagger{(\hat{\mathbf{H}}_{(j);n})}^\dagger\hat{\mathbf{G}}_{(k)}{(\hat{\mathbf{G}}_{(k)})}^\dagger\hat{\mathbf{H}}_{(j);n}\Bigg).\\
    \end{aligned}
     \label{equ:trace}
\end{equation}
The steps in~\eqref{equ:trace} can be explained as follows: (a) is due to the mixed-product property and the combination of the two summations; (b) is because basis matrices are orthogonal and are idempotent; (c) is due to the linearity of the trace operation and trace property of the Kronecker product; and (d) is due to the cyclic property of the trace \cite{van2000ubiquitous}.
To facilitate this derivation, \eqref{equ:trace} is first reformulated as follows:
\begin{equation}\small
    \begin{aligned}
        &\text{T}_{k,j} = \text{Tr}\bigg( \sum_{n=0}^{N-1}\bm{\mathcal{E}}_{n} \otimes  {{\mathbf{H}}^\prime_{n}}^\dagger \sum_{n=0}^{N-1}{\bm{\mathcal{E}}}_{n} \otimes  {\mathbf{H}}^\prime_{n}\bigg)\\
         &\quad=\sum_{n=0}^{N-1}\text{Tr}\bigg(( \mathbf{E}_{j}\otimes\bm{\mathcal{X}}_{(j)})\hat{\mathbf{H}}^\dagger_{(j);n}( \mathbf{E}_{k}\otimes\bm{\mathcal{X}}_{(k)})\hat{\mathbf{H}}_{(j);n}\bigg)\\
        &\quad=\sum_{n=0}^{N-1}\text{Tr}\bigg({\bigg[\hat{\mathbf{H}}_{(j);n}( \mathbf{E}_{j}\otimes\bm{\mathcal{X}}_{(j)})\bigg]}^\dagger( \mathbf{E}_{k}\otimes\bm{\mathcal{X}}_{(k)})\hat{\mathbf{H}}_{(j);n}\bigg)\\
    \end{aligned}
\end{equation}
where $\mathbf{E}_{i} $ (as defined in \textbf{Lemma 2}), was obtained through the use of the mixed product property on (d) from \eqref{equ:trace}. 
By definition, the trace can be expressed as follows:
\begin{equation}\small
    \begin{aligned}
        &\text{T}_{k,j} = \sum_{n=0}^{N-1}\text{Tr}\bigg({\bigg[\hat{\mathbf{H}}_{(j);n}( \mathbf{E}_{j}\otimes\bm{\mathcal{X}}_{(j)})\bigg]}^\dagger( \mathbf{E}_{k}\otimes\bm{\mathcal{X}}_{(k)})\hat{\mathbf{H}}_{(j);n}\bigg)\\
         &=\sum_{n=0}^{N-1}\sum_{m_1=0}^{\mathcal{R}_k-1}\sum_{m_2=0}^{\mathcal{R}_j-1}{\left|\hat{\mathbf{H}}_{(j);n}((k-1)M'+ m_1,(j-1)M' + m_2) \right|}^2.
    \label{equ:gentrace}
  \end{aligned}
\end{equation}
This can be used to compute both the desired signal strength and the interference experienced by the $k$-th UE, using $\text{T}_{k,k}$ and $\text{T}_{k,j}$  respectively. Finally, the SINR for OFDM is given by
\begin{equation}
\begin{aligned}
    \mathrm{SINR}_k = \frac{\text{T}_{k,k}}{\gamma^{-1} + \sum_{j=1,j\neq k}^\beta\text{T}_{k,j}}.
\end{aligned}
\label{eq:SINR_ofdm}
\end{equation}

\subsection{Upper Bound on SE for Delay-based Waveforms}
Here, we derive expressions for upper bounds on the \ac{SE} of each UE, which provides good insight into the average achieved SE performance.
Let $E_s$ and $N_0$ be the average symbol and noise powers, respectively. With the matrix definitions from Subsection~\ref{sec:MUI_delay_waveforms}, the achievable SE for the $k$-th UE can be written as \cite{tse2005fundamentals}
\begin{equation}
   \mathrm{\eta}_k = \frac{1}{K'}\log_2\det\left( \mathbf{I}_{K'} +\frac{E_s}{N_0 +\text{Tr}\big({(\mathbf{\Xi}_{(k)})}^\dagger\mathbf{\Xi}_{(k)}\big)} {(\mathbf{\Gamma}_{(k)})}^{\dagger}\mathbf{\Gamma}_{(k)} \right).
       \label{equ:spectral_definition}
\end{equation}
As discussed above, in this scenario the BS takes care of the overall excess delay in combination of $\lambda_k$ by having larger $g_k$. In this case, there is no interference from the $(k-1)$-th UE. However, this is still  problematic as it has a negative effect on the achievable \ac{SE} as will be shown next. 

The matrix ${(\mathbf{\Gamma}_{(k)})}^{\dagger}\mathbf{\Gamma}_{(k)}$ is positive semi-definite, and its eigenvalues are real and non-negative. Hence, the matrix $\mathbf{I}_{K'} +\frac{E_s}{N_0 +\text{Tr}\big({(\mathbf{\Xi}_{(k)})}^\dagger\mathbf{\Xi}_{(k)}\big)} {(\mathbf{\Gamma}_{(k)})}^{\dagger}\mathbf{\Gamma}_{(k)}$ is positive-definite. Let the eigenvalues of
${(\mathbf{\Gamma}_{(k)})}^{\dagger}{\mathbf{\Gamma}_{(k)}}$ be $\aleph_{k,1},\aleph_{k,2},...,\aleph_{k,K'}$, we can express the \ac{SE} of the $k$-th UE as 
\begin{equation}
\begin{aligned}
     \mathrm{\eta}_k &= \frac{1}{K'}\log_2\bigg[
   \prod_{i=1}^{K'}\bigg(
   1 +\frac{\aleph_{k,i}}{\gamma^{-1} +{\text{Tr}\left({(\mathbf{\Xi}_{(k)})}^\dagger\mathbf{\Xi}_{(k)}\right)}}\bigg) \bigg]\\
   &=\frac{1}{K'}\sum_{i=1}^{K'} \bigg(\log_2\bigg[
   1 +\frac{\aleph_{k,i}}{\gamma^{-1} +{\text{Tr}\left({(\mathbf{\Xi}_{(k)})}^\dagger\mathbf{\Xi}_{(k)}\right)}}\bigg] \bigg).\\
\end{aligned}
\end{equation}
Using Jensen's inequality, we can give an upper bound on the achievable SE of the $k$-th UE. The inequality states that the mean of a concave function (such as $\text{log}_2(1+x)$) of a random variable is upper-bounded by the function evaluated at the mean of the random variable.  With this, an upper bound for the achievable SE is given by
\begin{equation}
   \mathrm{\eta}_k \leq \mathrm{\eta}^{\text{bound}}_k = 
   \log_2\left( 1 +\frac{1}{K'}\frac{\text{Tr}( {(\mathbf{\Gamma}_{(k)})}^{\dagger}{\mathbf{\Gamma}_{(k)}})}{\gamma^{-1} +{\text{Tr}\left({(\mathbf{\Xi}_{(k)})}^\dagger\mathbf{\Xi}_{(k)}\right)}} \right),
\end{equation}
where the trace is used to express the sum of the eigenvalues. By definition, we have
\begin{equation}
    \mathrm{\eta}^{\text{bound}}_k = 
  \log_2\left( 1 +\frac{\sigma_k}{\gamma^{-1} +K'{\zeta_k}} \right).
\end{equation}
Hence, both $\sigma_k$ and $\zeta_k$, are important to consider. While using guard reduces the interference $\zeta_k$, it also reduces the signal strength $\sigma_k$, thereby not achieving the optimal SE available. 

\subsection{Upper Bound on SE for OFDM Waveform}
Using arguments similar to the above, we can show that the upper bound in the case of OFDM is given by 

\begin{equation}
    \mathrm{\eta}^{\text{bound}}_k = 
  \log_2\left( 1 +\frac{\text{T}_{k,k}}{\gamma^{-1} + \sum_{j=1,j\neq k}^\beta\text{T}_{k,j}} \right).
\end{equation}

\subsection{\textcolor{black}{Impact of Doppler Spread}}
\label{sec:doppler_for_each_waveform}
\textcolor{black}{
The above derivations rely on the key assumption of no Doppler, which enables a unified treatment across the considered waveforms. Once Doppler is included, waveform-specific differences in the input--output relations become unavoidable. We have already seen one example in Section~\ref{sec:including_doppler_for_ofdm}, where OFDM was treated separately. Similarly, starting from the time-delay channel with Doppler in \eqref{equ:Ht}, the OTFS input--output relation in the delay-Doppler domain can be written as \cite{raviteja2018practical}:
\begin{equation}
\mathbf{H}_{(u)}^\mathrm{OTFS}
=
(\mathbf{F}_N\otimes \mathbf{I}_M)
\mathbf{H}_{(u)}^\mathrm{TD}
(\mathbf{F}^\dagger_N \otimes \mathbf{I}_M).
\end{equation}
Similarly, OTSM is represented by \cite{otsm_transaction}:
\begin{equation}
\mathbf{H}_{(u)}^\mathrm{OTSM}
=
(\mathbf{W}_N\otimes \mathbf{I}_M)
\mathbf{H}_{(u)}^\mathrm{TD}
(\mathbf{W}^\dagger_N \otimes \mathbf{I}_M),
\end{equation}
where \(\mathbf{W}_N\) is the $N \times N$ Hadamard transform matrix.
For ODDM, whose operating domain is the delay-Doppler domain, the effective channel is given by \cite{tong2024orthogonal}:
\begin{equation}
\mathbf{H}_{(u)}^\mathrm{ODDM}
=
(\mathbf{I}_M\otimes \mathbf{F}_N)
\mathbf{\Pi}^\dagger
\mathbf{H}_{(u)}^\mathrm{TD}
\mathbf{\Pi}
(\mathbf{I}_M \otimes \mathbf{F}^\dagger_N),
\end{equation}
where \(\mathbf{\Pi}\) is a permutation matrix for staggering ODDM symbols in time \cite{tong2024orthogonal}. For OCDM, whose operating domain is the Fresnel domain, the effective channel is given by \cite{haif2024novel}:
\begin{equation}
\mathbf{H}_{(u)}^\mathrm{OCDM}
=
(\mathbf{I}_N\otimes \mathbf{\Phi}_M)
\mathbf{H}_{(u)}^\mathrm{TD}
(\mathbf{I}_N \otimes \mathbf{\Phi}^\dagger_M),
\end{equation}
where \(\mathbf{\Phi}_M\) is the discrete Fresnel transform matrix. Although these expressions do not lead to a unified analysis, they allow each waveform to be evaluated under the more realistic scenario of Doppler spread.
}

\begin{figure}[!t]
\centering
\includegraphics[scale = .45]{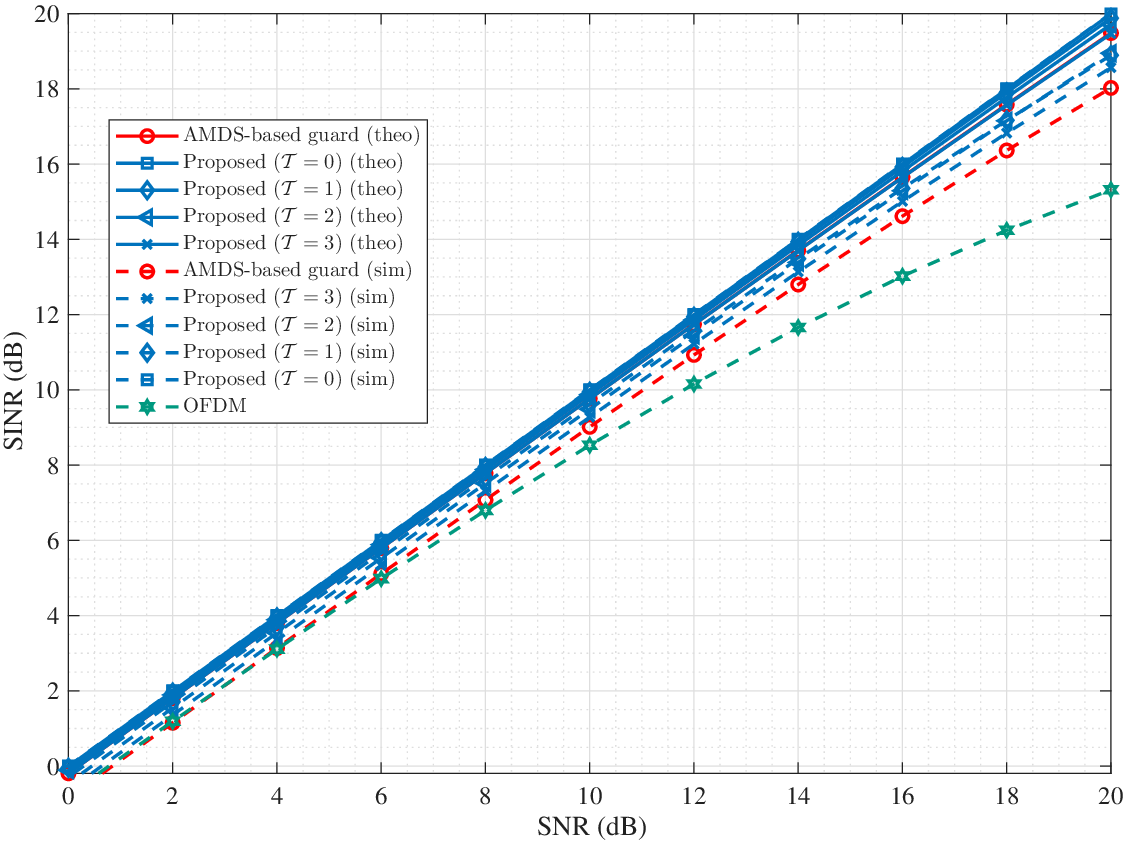}
\caption{The average received SINR at the BS for different schemes.}
\label{fig:sinr_average}
\end{figure}

\textcolor{black}{\subsection{Exceeding the CP length}
\label{sec:exceeding_the_cp}
In addition to the no-Doppler assumption, the preceding derivations rely on the block-circular structure induced by the CP. In particular, they assume that the combined channel delay spread and residual TM remain within the CP duration, so that no inter-block leakage occurs. When this condition is violated, the effective channel is no longer block diagonal, and additional \ac{ISI} and \ac{ICI} terms appear.\\
Although practical \ac{TA} is expected to keep the residual misalignment within the CP duration, it is useful to examine this beyond-CP case as a stress-test scenario. This allows us to evaluate how the considered waveforms behave when the within-CP assumption is violated.\\
The modeling can be performed as follows. We start by including CP insertion using the matrix:
\begin{equation}
     \mathbf{A}_{\rm CP}= \mathbf{I}_{N}\otimes\begin{bmatrix}
 \mathbf{0}_{L_{\rm CP}\times (M-L_{\rm CP})}~~~~ \mathbf{I}_{L_{\rm CP}}  \\
\mathbf{I}_M \\
\end{bmatrix}.
     \label{eq:CP_add_matrix}
\end{equation}
Similarly, let $\mathbf{R}_{\rm CP} =\mathbf{I}_{N}\otimes[ \mathbf{0}_{M \times L_{\rm CP}}~\mathbf{I}_M]$ be the CP removal matrix. This models CP addition and removal for each symbol in the frame. We can also model inter-frame interference by including the part of the previous frame within the TM value, or we can assume the guards between frames are made intentionally large enough that the TM does not exceed them. Even with this simplifying assumption (not accounting for inter-frame interference), the time-delay channel modeling changes and the IOR becomes
\begin{equation}
    \mathbf{H}^{\rm CP}_{(k)} = \mathbf{R}_{\rm CP}\bigg(\sum_{i=1}^{P_k} h_{k,i} \boldsymbol{\Pi}_{(M+L_{\rm CP})N}^{l_{k,i}+\lambda_k}\bigg)\mathbf{A}_{\rm CP},
    \label{eq:CP_issue}
\end{equation}
which does not have the block-diagonal structure we have assumed throughout the analysis. However, we can still assess the effect through numerical evaluations as will be seen in Section~\ref{sec:simulation}.}

\begin{figure}[!t]
\centering
\includegraphics[scale = .45]{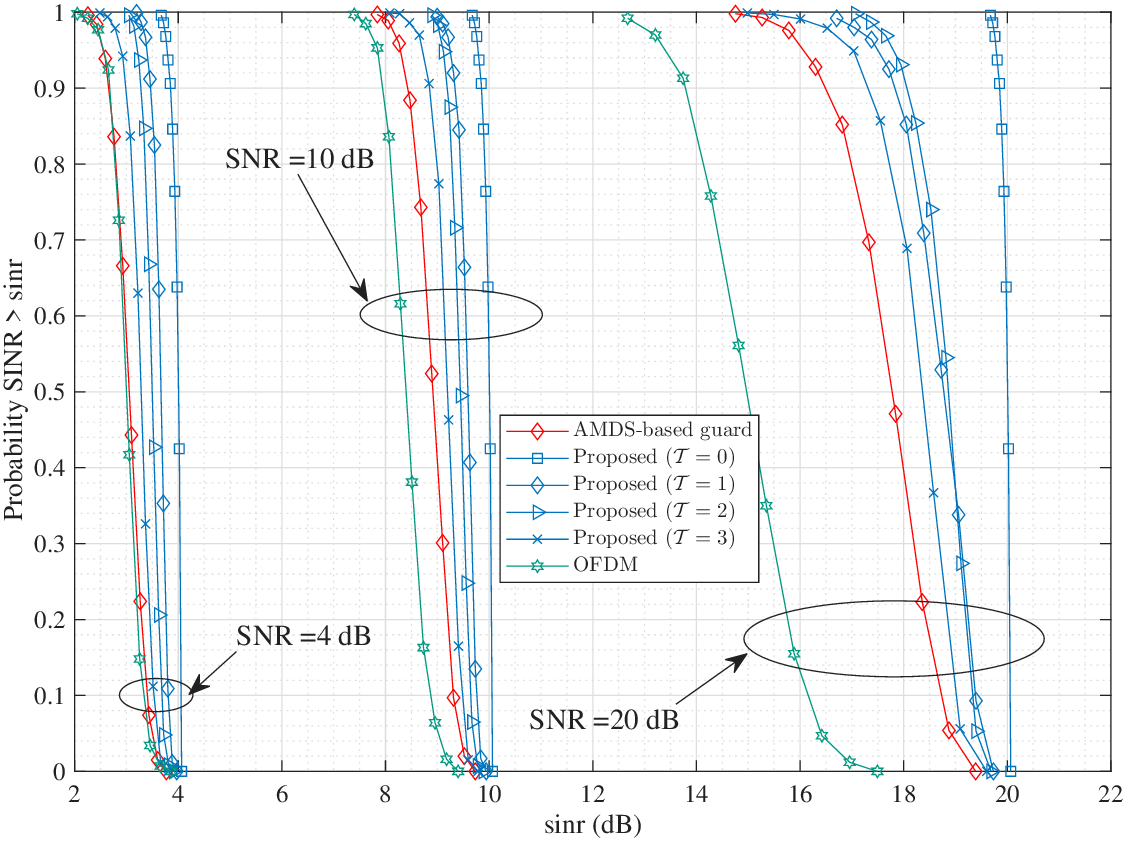}
\caption{CCDF of the average received SINR at the BS for different schemes.}
\label{fig:sinr_ccdf}
\end{figure}

\section{Simulation Results}
\label{sec:simulation}
This section evaluates the performance of the proposed scheme and validates the analytical results through simulations. We consider an uplink scenario involving $K_u = 4$ UEs scheduled along the delay domain. As a baseline, we set $M' = 16$ and $l_{k,\max} = l_{\max} =4$ for all UEs.

\subsection{TM impact on the general delay-based IOR}

We explore both alternatives for guard allocation and their effects on the SINR and SE (i.e., whether the guards are allocated based on the MMDS or AMDS, in the presence of different propagation delays). For OFDM simulations, the same number of UEs are scheduled along the frequency domain. As discussed earlier, we consider different possible estimation errors in the estimated $\tilde{\lambda}_k$ values. 
Throughout this section, we compare the proposed scheme, referred to as ``Proposed,'' for different $\mathcal{T}$ values (i.e., different estimation-error levels), and the conventional scheme, labeled ``AMDS-based guard'' or ``MMDS-based guard'' \, corresponding to the two alternatives of guard allocation. The analytical results derived based on the simplifying assumptions of ignoring fractional delays (labeled ``theo'' in the figures) are compared against a realistic scenario involving fractional delays (labeled ``sim'' in the figures). In this subsection, we isolate the impact of \ac{TM} on the unified delay-based \ac{IOR} by neglecting Doppler spread for the delay-based waveforms, while OFDM is evaluated under Doppler spread as the reference impairment. The joint impact of Doppler spread and \ac{TM} on each individual waveform is evaluated separately in Subsection~\ref{sec:sim_joint_doppler_tm}. \textcolor{black}{The misalignment size $\lambda_k$ is generated from a uniform distribution $\lambda_k\sim \mathcal{U}(0,4)$ including fractional values. The upper value $4$ is chosen to produce noticeable timing-induced delay-domain shifts while keeping the dominant leakage mainly between adjacent user allocations, which is consistent with the adjacent-leakage assumption used in the analysis. In addition, since the baseline maximum channel delay spread is set to $l_{\max}=4$, this choice makes the residual timing offset comparable to the physical channel delay spread. Therefore, the adopted TM setting represents a meaningful stress case for the considered delay-domain allocation while remaining consistent with the assumptions of the analytical model.} 

\begin{figure}[!t]
\centering
\includegraphics[scale = .45]{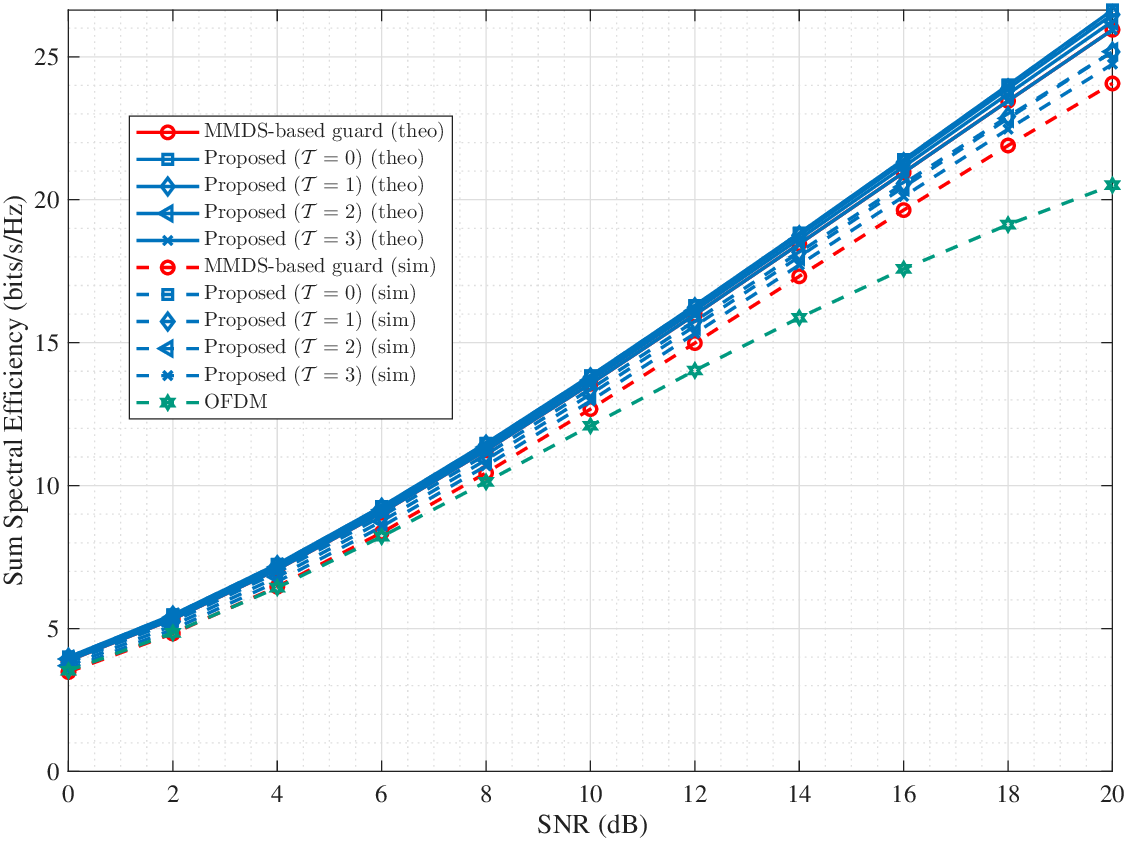}
\caption{The average received sum SE at the BS for different schemes.}
\label{fig:se_average}
\end{figure}

\subsubsection{Multi-User Interference}

In this setup, the BS allocates delay guards based on the aligned MDS (AMDS) of the UEs' channels, without taking into consideration the nonzero $\lambda_k$ values.\\
{
Fig.~\ref{fig:sinr_average} shows the average received SINR according to the setup described above. First, we observe that the analytical prediction closely follows the simulated performance. Although the gaps vary across the considered TM estimation accuracies, they remain small, not exceeding 2~dB, supporting the accuracy of the analysis and enabling fair comparisons among the schemes. For the proposed scheme, as the TM estimation becomes more accurate (i.e., lower $\mathcal{T}$ values) the SINR is improved. As $\mathcal{T}$ increases, the SINR performance of the proposed system degrades, but it keeps a net gain even in the worst case of $\mathcal{T}=3$. While OFDM is not affected by TM, Doppler spread significantly degrades the SINR performance, which starts to flatten at high SNR.\\
To provide a more comprehensive view of the SINR performance, we also include the \ac{CCDF} of the achieved SINR for the different schemes,  Fig.~\ref{fig:sinr_ccdf}. In addition to the improved average SINR performance, we also notice that the conventional scheme exhibits a more dispersed SINR distribution (i.e., SINR is likely to be lower than its average). As shown, OFDM under Doppler yields the most dispersed SINR distribution, indicating that it is likely to fall below its average performance. This leads to a degradation in the upper bounds for SE as shown next.
}

% \textcolor{gray}{}

\begin{figure}[!t]
\centering
\includegraphics[scale = .45]{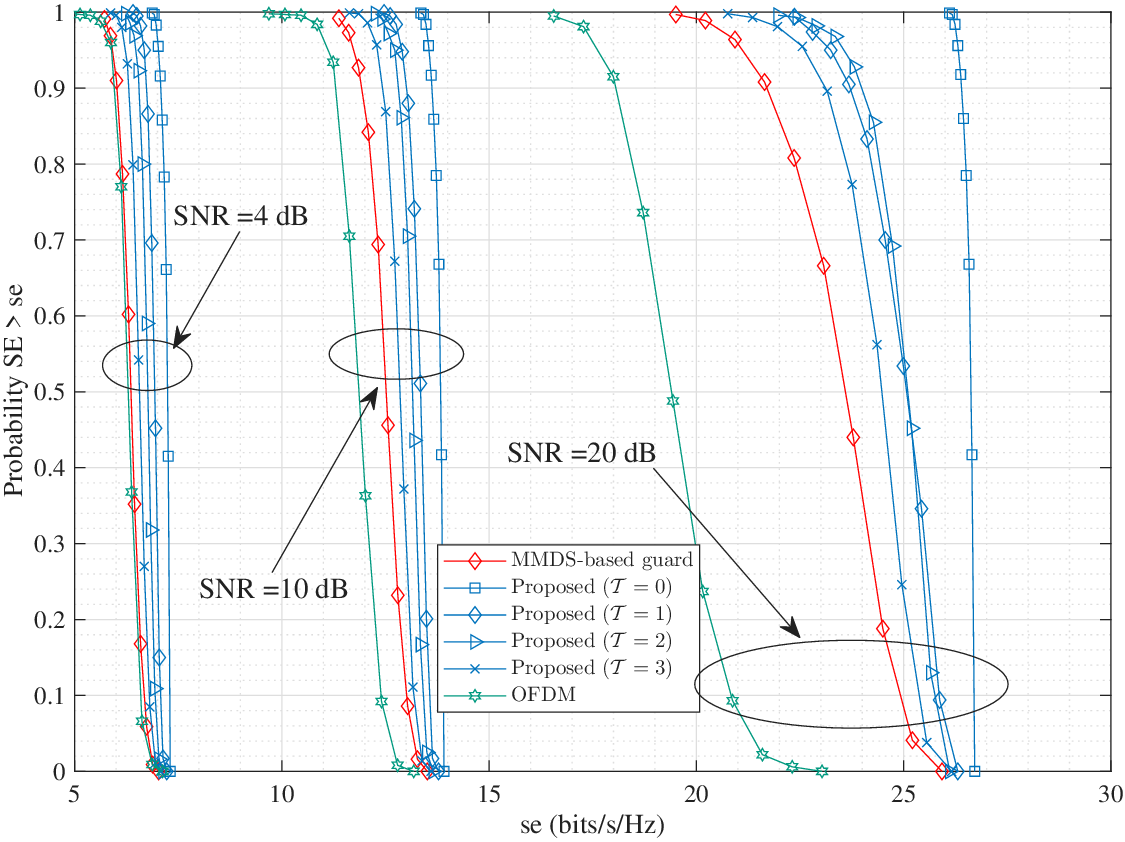}
\caption{CCDF of the average received sum SE at the BS for different schemes.}
\label{fig:se_ccdf}
\end{figure}

\subsubsection{Spectral Efficiency}

In this setup, the BS allocates delay guards for the $k$-th UE based on the total \ac{MMDS} of the corresponding channel, which exceeds the \ac{AMDS} by $\lambda_k$, thereby reducing the efficiency of the system.

Fig.~\ref{fig:se_average} presents the sum SE received at the BS, for OFDM and delay-based waveforms. The theoretical upper bounds (ignoring fractional delay, labeled ``theo'' in the figures) are compared against the simulated upper bounds (accounting for fractional values, labeled ``sim'' in the figures). Similar to SINR, we notice agreement between the analytically predicted behavior and the obtained performance through simulation. The maximum gap is about 3 bits/s/Hz. First, we observe that the biggest difference between OFDM and delay-based waveforms is that the spectral efficiency of the former flattens quickly in the presence of Doppler. On the other hand, as the TM estimation becomes more accurate, i.e., for lower $\mathcal{T}$ values, the simulated sum SE is increased, and the gap with the theoretical curves becomes tighter.  As $\mathcal{T}$ increases, the SE performance of the proposed system degrades. Compared to delay-based schemes, OFDM exhibits degraded SE performance, and the gap increases with SNR. 

We additionally show the CCDF of the achieved SE according to the different schemes, following the same methodology used for the SINR performance, Fig.~\ref{fig:se_ccdf}. Once more, we see that the conventional scheme shows a more dispersed distribution, especially at high SNR, while OFDM exhibits the most dispersed SE distribution, indicating that it can reach values much lower than average. 

Finally, we offer some intuition for the observed performance gain of delay-based waveforms compared with OFDM in the considered setup. While delay and Doppler may seem mathematically symmetric when their specific magnitudes are ignored, delay-based waveforms can benefit from a useful physical characteristic of the adopted channel model. Under the assumed power delay profile, paths with larger delays have smaller average powers. Consequently, the leakage caused by an interfering user becomes less severe farther away in the delay domain. With the addition of guard resources, such interference can be suppressed more effectively. By contrast, the interference experienced by OFDM due to Doppler does not enjoy a similar effect, since there is generally no physical reason for paths with larger Doppler shifts to have smaller gains. This provides one possible interpretation of the performance advantage observed for delay-based waveforms over OFDM in this setup. 

\begin{figure}[!t]
\centering
\includegraphics[scale = .45]{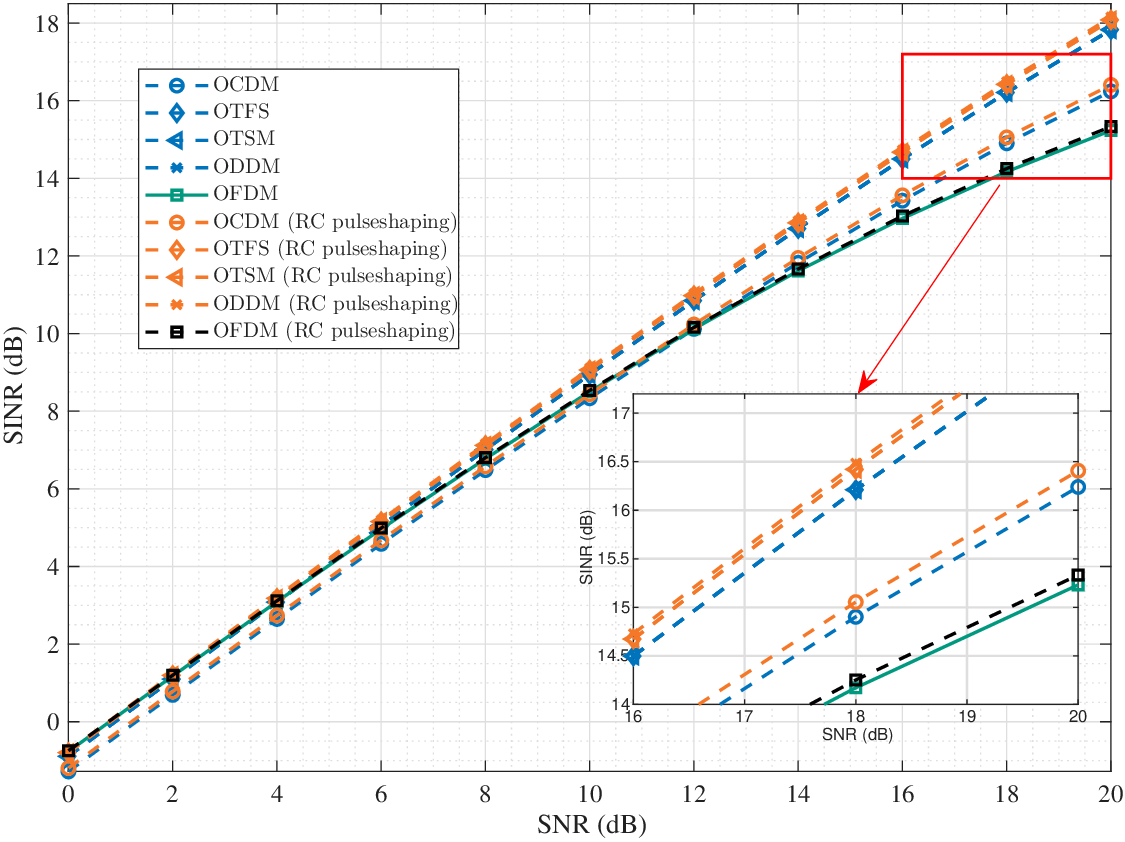}
\caption{SINR evaluation including Doppler for each waveform.}
\label{fig:sinr_doppler_each_waveform}
\end{figure}

\textcolor{black}{\subsection{Combined Impact of Doppler and TM on Individual Waveforms}
\label{sec:sim_joint_doppler_tm}
As discussed in Section~\ref{sec:doppler_for_each_waveform}, we can assess the performance of each waveform individually while including Doppler spread.
\subsubsection{Combined impact on SINR performance}
We first assess the effect of Doppler in addition to TM on the SINR performance for each waveform, Fig.~\ref{fig:sinr_doppler_each_waveform}. The blue curves represent the performance of the delay-based waveforms while the green curve is for OFDM. As can be seen, including Doppler for OTFS, ODDM, and OTSM results in negligible SINR loss, because \ac{DeDMA} is used. In the considered delay-division scheme (which can be adopted for OTFS, OTSM, and ODDM), Doppler spread causes negligible interference among users (although it still has to be estimated and equalized at the receiver). This illustrates the advantage of delay division: even though it struggles with TM, it offers strong robustness against Doppler. This is not the case for OFDM and OCDM, where the operating domains are the frequency and Fresnel domains, respectively. For these two waveforms, Doppler spread causes inter-user interference and degrades the SINR. Finally, we note that OFDM suffers more SINR degradation than OCDM, with OCDM outperforming OFDM in the presence of TM and Doppler.\\
We also include the effect of pulse shaping in mitigating fractional leakage, as shown by the orange and black curves in Fig.~\ref{fig:sinr_doppler_each_waveform}. \Ac{RC} pulse shaping is used with a roll-off factor of 0.2. The results show that pulse shaping can be used to effectively mitigate interference across all waveforms, while the relative performance is maintained.
}

\begin{figure}[!t]
\centering
\includegraphics[scale = .45]{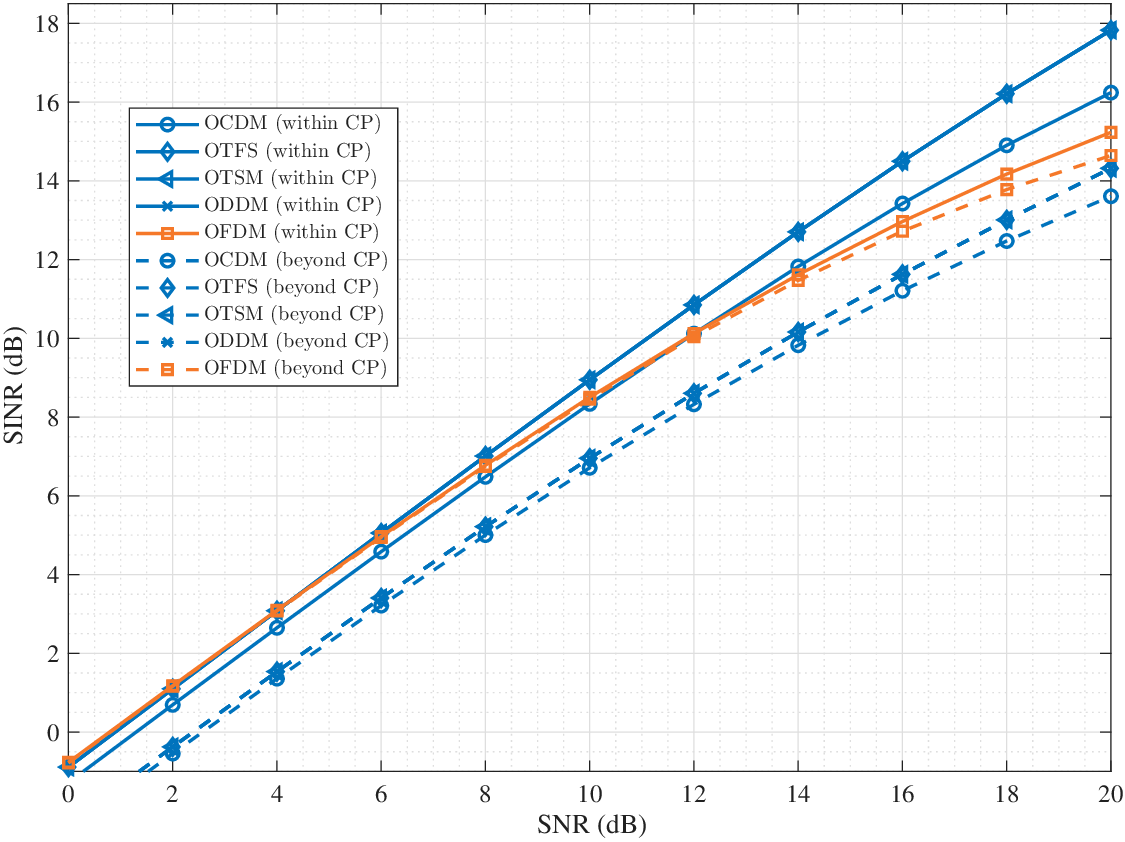}
\caption{\textcolor{black}{Effect of a combined delay spread exceeding the CP length on SINR performance.}}
\label{fig:beyond_CP}
\end{figure}

\textcolor{black}{\subsubsection{Exceeding the CP length}
Furthermore, as discussed in Section~\ref{sec:exceeding_the_cp}, we assess the case where misalignment pushes the effective delay spread beyond the CP length, as shown in Fig.~\ref{fig:beyond_CP}. In this scenario, the CP length \(L_{\rm CP}\) exactly covers the channel delay spread. The TM is generated from \(\mathcal{U}(0,L_{\rm CP})\), such that the total effective spread can reach twice the CP length. First, we observe that even OFDM degrades under this level of misalignment, since the loss of circularity destroys orthogonality and results in ISI and ICI. In addition, the inter-user leakage discussed earlier increases as the TM magnitude increases. As a result, all waveforms experience performance degradation, with delay-based waveforms experiencing the largest degradation. This effect can be mitigated or avoided by several approaches. For example, OTFS can benefit from a reduced-CP structure, where the CP length can be increased without introducing excessive overhead, unlike full-CP OFDM, which cannot afford to enlarge the CP as much. In practice, TA is expected to keep the combined channel delay spread and residual TM within the CP duration; therefore, this scenario should be interpreted as a stress-test case that illustrates the behavior of the considered waveforms when the within-CP assumption is violated.
}

\begin{figure}[!t]
\centering
\includegraphics[scale = .45]{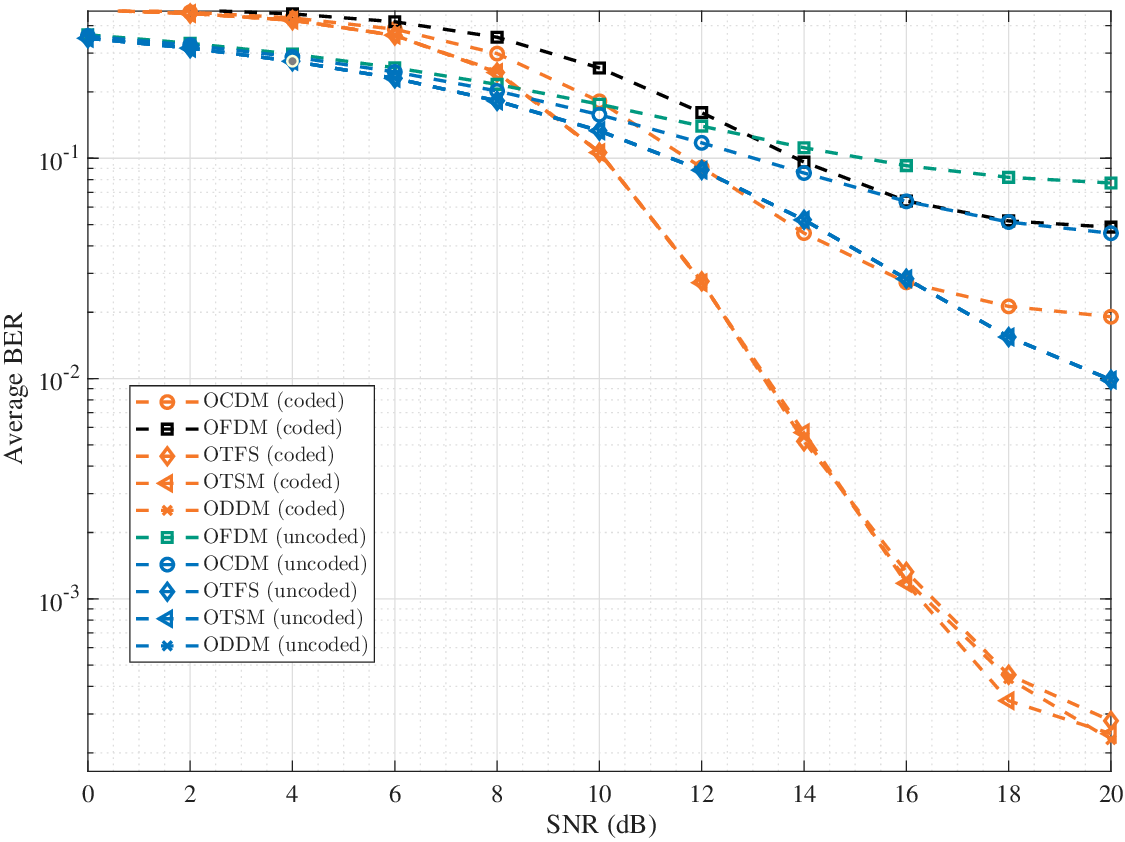}
\caption{\textcolor{black}{BER performance for all waveforms with and without coding.}}
\label{fig:BER}
\end{figure}

\textcolor{black}{\subsubsection{BER performance}
Finally, we assess the BER performance for each waveform under both Doppler and TM. Using the different effective channel representations (Section~\ref{sec:doppler_for_each_waveform}), we can assess the detection and equalization performance for each waveform. We then include coding for each waveform. The results are shown in Fig.~\ref{fig:BER}, for \ac{QPSK} modulation and coding rate 1/2. As can be seen from the results, coding can significantly enhance the performance and mitigate the interference caused by TM and Doppler for all waveforms. However, it is also noticeable that ODDM/OTSM/OTFS benefit from coding to a much greater extent, achieving interference-robust performance, followed by OCDM and then OFDM.
}

\section{Conclusion} % 6G applications and requirements
\textcolor{black}{This work examined the problem of timing misalignment (TM) in multi-user uplink transmission for a class of delay-based waveforms, namely OTFS, OTSM, OCDM, and ODDM. Unlike prior studies that typically analyze TM in a waveform-specific manner, this work developed a unified input--output formulation that reveals a common structural effect of residual propagation delay across these waveforms. In particular, we showed that a residual misalignment, even within the CP length, translates into actual shifts in the delay domain. As a result, TM gives rise to a fundamental trade-off: it either increases MUI or reduces spectral efficiency through the need for larger guard allocations. In contrast, OFDM is unaffected by TM under the considered within-CP condition, since timing offsets manifest only as phase rotations. Rather, OFDM's main bottleneck in the considered benchmark is the interference caused by Doppler spread.\\
Beyond the unified Doppler-free analysis, we further evaluated the waveform-specific impact of joint Doppler spread and \ac{TM}, the case where the combined delay spread exceeds the CP duration, and coded BER performance. The results show that delay-based waveforms remain attractive under Doppler, but they require careful handling of residual timing offsets. These findings motivate future work on practical TM estimation, fractional-delay compensation, receiver-side interference cancellation, and joint scheduling strategies for delay-based multi-user uplink systems.}

\appendices

%% I believe we only need a  mod M for (i+l)mod M and (i+l')mod M %%%%%%%%%%%%%%%%%%%%%%%%%%%%%%%%%%%%%%%%%%%%%%%%%%%%%%%%%%%%%%%%%%%%%%%%%%%%%%%%%%%%

\section{Proof of Lemma 3}
The proof is found by applying the trace definition
\begin{equation}
\begin{aligned}
    \text{Tr}({\bm{\mathcal{B}}_{{(k-1)}}}^{\dagger}{\bm{\mathcal{A}}_{(k)}}) 
    =\sum_{j=1}^{M}\sum_{i=1}^{M}\bm{\mathcal{B}}_{{(k-1)}}(i,j)\bm{\mathcal{A}}_{(k)}(i,j),
\end{aligned}
\end{equation} with the help of \textbf{Lemma 2} and \eqref{eq:cor}.\hfill$\blacksquare$

\section{Proof of Lemma 4}
The proof is, once more, found by applying the trace~definition
\begin{equation}
\begin{aligned}
    \text{Tr}({\bm{\mathcal{B}}'_{(k)}}^{\dagger}{\bm{\mathcal{A}}'_{(k)}}) 
    =\sum_{j=1}^{M}\sum_{i=1}^{M}\bm{\mathcal{B}}'_{(k)}(i,j)\bm{\mathcal{A}}'_{(k)}(i,j),
\end{aligned}
\end{equation} and \textbf{Lemma 2} in combination with \eqref{eq:cor1} and \eqref{eq:cor2}.\hfill$\blacksquare$

\section{Proof of Theorem 1}
By expanding \eqref{equ:hdhdh} we find
\begin{equation}\small
\begin{aligned}
    \zeta_k=& \frac{1}{K'}\text{Tr}\big(\mathbf{G}^\dagger_{(k-1)}\mathbf{Q}^\dagger_{(k-1)}\mathbf{H}^\dagger_{(k-1)}\mathbf{Q}_{(k)}\mathbf{Q}^\dagger_{(k)}\mathbf{H}_{(k-1)}\mathbf{Q}_{(k-1)}\mathbf{G}_{(k-1)}\big) \\
    =& \frac{1}{K'}\text{Tr}\big(\mathbf{Q}_{(k-1)}\mathbf{G}_{(k-1)}\mathbf{G}^\dagger_{(k-1)}\mathbf{Q}^\dagger_{(k-1)}\mathbf{H}^\dagger_{(k-1)}\mathbf{Q}_{(k)}\mathbf{Q}^\dagger_{(k)}\mathbf{H}_{(k-1)}\big), 
\end{aligned}
\end{equation}
where we have used the trace circular property; next we use the trace of a Kronecker, which is the multiplication of the traces \cite{van2000ubiquitous} with $\text{Tr}(\mathbf{I}_{N}) = N$.
In other words,
\begin{equation}\small
\begin{aligned}
    &\zeta_k\\
    &=\frac{N}{K'}\text{Tr}\big(\bm{\mathcal{B}}_{(k-1)}\sum_{j=1}^{P_{k-1}} h^*_{k-1,j} \bm{\Pi}^{-l'_{k-1,j}}_{M}\bm{\mathcal{A}}_{(k)}\sum_{i=1}^{P_{k-1}} h_{k-1,i} \bm{\Pi}^{l'_{k-1,i}}_{M}\big) \\
    ~~&= \frac{N}{K'}\sum_{j=1}^{P_{k-1}}\sum_{i=1}^{P_{k-1}}h^*_{k-1,j} h_{k-1,i}\text{Tr}\big(\bm{\mathcal{B}}_{(k-1)}  \bm{\Pi}^{-l'_{k-1,j}}_{M}\bm{\mathcal{A}}_{(k)} \bm{\Pi}^{l'_{k-1,i}}_{M}\big) \\
    ~~&= \frac{N}{K'}\sum_{j=1}^{P_{k-1}}\sum_{i=1}^{P_{k-1}}h^*_{k-1,j} h_{k-1,i}\text{Tr}\big({(\bm{\Pi}^{l'_{k-1,j}}_{M}\bm{\mathcal{B}}_{(k-1)}  )}^\dagger\bm{\mathcal{A}}_{(k)} \bm{\Pi}^{l'_{k-1,i}}_{M}\big), \\
\end{aligned}
\end{equation}
where $ l'_{k-1,j}=l_{k-1,j}+\delta_{k-1}$.
Now, we invoke \textbf{Lemma 3} to find that
\begin{equation}
\begin{aligned}
    &\zeta_k= \frac{N}{K'}\sum_{j=1}^{P_{k-1}}\sum_{i=1}^{P_{k-1}}h^*_{k-1,j} h_{k-1,i}\text{Tr}\big({(\bm{\Pi}^{l'_{k-1,j}}_{M}\bm{\mathcal{B}}_{(k-1)}  )}^\dagger\\
    &\times\bm{\mathcal{A}}_{(k)} \bm{\Pi}^{l'_{k-1,i}}_{M}\big) \\
      ~~&= \frac{N}{K'}\sum_{i=1}^{P_{k-1}} {|h_{k-1,i}|}^2r[l_{k-1,i}+\delta_{k-1}-g_{k-1}]. 
\end{aligned}
\end{equation}
This completes the proof.\hfill$\blacksquare$

\section*{Acknowledgment}
Funded by the European Union under the Grant Agreement No 101293106. Views and opinions expressed are however those of the author(s) only and do not necessarily reflect those of the European Union or Smart Networks and Services Joint Undertaking. Neither the European Union nor the granting authority can be held responsible for them.

\bibliographystyle{IEEEtran}
\bibliography{IEEEfull}
\vfill
\end{document}